\documentclass{siamltex}

\usepackage{latexsym}
\usepackage{epsfig}
\usepackage{cite}
\usepackage{amssymb}
\usepackage{amsmath}
\usepackage{float}
\usepackage[usenames]{color}

\newcommand{\cA}{\mathcal{A}}
\newcommand{\D}{\mathcal{D}}
\newcommand{\cE}{\mathcal{E}}
\newcommand{\cG}{\mathcal{G}}
\newcommand{\cI}{\mathcal{I}}
\newcommand{\cU}{\mathcal{U}}
\newcommand{\cV}{\mathcal{V}}
\newcommand{\cS}{\mathcal{S}}
\newcommand{\cX}{\mathcal{X}}
\newcommand{\cY}{\mathcal{Y}}

\newcommand{\R}{\mathbb R}

\newcommand{\eps}{\varepsilon}

\newcommand{\EX}{\hbox{\bf E}}
\newcommand{\pred}{\mathcal{B}}
\newcommand{\etal}{\emph{et al.}}
\newcommand{\tsearch}{\mathcal{B}} 
\newcommand{\tconf}{\mathcal{C}} 
\newcommand{\eqdef}{:=}

\floatstyle{ruled}
\newfloat{algorithm}{thp}{alg}
\floatname{algorithm}{Algorithm}

\newtheorem{claim}[theorem]{Claim}

\title{Self-Improving 
Algorithms\thanks{Preliminary versions appeared as
N. Ailon, B. Chazelle, S. Comandur, and D. Liu,
\emph{Self-improving Algorithms} in Proc.~17th SODA, pp.~261--270, 2006;
and K. L. Clarkson and C. Seshadhri, \emph{Self-improving Algorithms for
Delaunay Triangulations} in Proc.~24th SoCG, pp.~148--155, 2008.
This work was supported
in part by NSF grants CCR-998817, 0306283, 
ARO Grant DAAH04-96-1-0181.}}

\author{
  {Nir Ailon}\thanks{Computer Science Faculty, Technion, Haifa, Israel} 
\and
  {Bernard Chazelle}\thanks{Department of Computer Science,
       Princeton University, Princeton, NJ, USA}
\and
  {Kenneth L. Clarkson}\thanks{IBM Almaden Research Center, San Jose, CA, USA}
\and
  {Ding Liu}\footnotemark[3]
\and
  {Wolfgang Mulzer}\thanks{Institut f\"ur Informatik,
     Freie Universit\"at Berlin, 14195 Berlin, Germany}
\and
  {C. Seshadhri}\footnotemark[4]
}

\begin{document}

\maketitle

\begin{abstract}
We investigate ways in which an algorithm can improve
its expected performance by fine-tuning itself
automatically with respect to an \emph{unknown} input 
distribution $\D$.
We assume here that $\D$ is of \emph{product type}.
More precisely, suppose that we need to process a sequence
$I_1, I_2, \ldots$ of inputs $I = (x_1, x_2, \ldots, x_n)$ of
some fixed length $n$, where each $x_i$ is drawn independently from some
\emph{arbitrary, unknown} distribution $\D_i$. The goal is to
design an algorithm for these inputs so that eventually
the expected running time will be optimal for the 
input distribution $\D = \prod_i \D_i$.

We give such \emph{self-improving} algorithms  
for two problems: (i) sorting a sequence of numbers and (ii) computing 
the Delaunay triangulation of a planar point set. 
Both algorithms achieve optimal expected limiting complexity.
The algorithms begin with a training phase
during which they collect information about the input distribution,
followed by a stationary regime in which the algorithms settle to their
optimized incarnations. 
\end{abstract}



\pagestyle{myheadings}
\thispagestyle{plain}
\markboth{AILON ET AL.}{SELF-IMPROVING ALGORITHMS}

\section{Introduction}\label{sec:introduction}

The classical approach to analyzing algorithms 
draws a familiar litany of complaints:
worst-case bounds are too pessimistic
in practice, say the critics, while
average-case complexity too often rests on unrealistic assumptions.
The charges are not without merit.
Hard as it is to argue that the only permutations we ever
want to sort are random, it is a different level of implausibility
altogether to pretend that the sites of a Voronoi diagram 
should always follow a Poisson
process or that ray tracing in a BSP tree should be spawned by
a Gaussian. Efforts have been made 
to analyze algorithms under more complex models
(eg, Gaussian mixtures, Markov model outputs) but 
with limited success and lingering doubts about 
the choice of priors.

Suppose we wish to compute a function $f$ that takes
$I$ as input. We get a sequence of inputs $I_1,I_2,\ldots$, and
wish to compute $f(I_1)$, $f(I_2),\ldots$.
It is quite plausible to assume that all these inputs
are somehow related to each other.
This relationship, though exploitable, may be very difficult
to express concisely.
One way of modeling this situation is to postulate
%imagine there being 
a fixed (but complicated) unknown distribution $\D$ of inputs.
Each input $I_j$ is chosen independently at random from $\D$.
Is it possible to learn quickly something about $\D$
so that we can compute $f(I)$ ($I$ chosen from $\D$) faster?
(Naturally, this is by no means the only possible input model. 
For	 example, we could have
a memoryless Markov source, where each $I_j$ depends only
on $I_{j-1}$. However, for simplicity we will
here focus on a fixed source that generates the
inputs independently.)

That is what a \emph{self-improving algorithm} attempts to do.
Initially, since nothing is know about $\D$, our 
self-improving algorithm can only provide some worst-case guarantee. 
As the algorithm sees more and more inputs, it can learn something about
the structure of $\D$. We call this
the \emph{training phase} of the self-improving algorithm.
During this phase, the algorithm collects and organizes information
about the inputs in the hope that it can be used
to improve the running time (with respect to inputs
from $\D)$. The algorithm then moves to the \emph{limiting phase}.
Having decided that enough has been learned about $\D$,
the algorithm uses this information to compute $f(I)$
faster. Note that this behavior is tuned to the distribution $\D$.
\begin{figure}
\begin{center}
\includegraphics{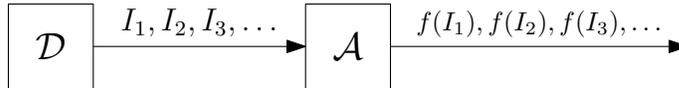}
\end{center}
\caption{A self-improving algorithm $\cA$ processes a sequence 
$I_1, I_2, \ldots$ of inputs drawn independently from a random source
$\D$.}
\end{figure}

Obviously, there is no reason why we should get a faster running time
for all $\D$. Indeed, if $f$ is the sorting function and $\D$ is the 
uniform distribution over permutations, then we require expected 
$\Omega(n \log n)$ time to sort. 
On the other hand, if $\D$ was a low-entropy source of inputs, 
it is quite reasonable to hope for a faster algorithm. 
So when can we improve our running time? An elegant way of expressing
this is to associate (using information theory) an ``optimal"
running time for each distribution. This is a sort of estimate
of the best expected running time we can hope for, given
inputs chosen from a fixed distribution $\D$. Naturally, the lower
the entropy of $\D$, the lower this running time will be.
In the limiting phase, our self-improving algorithm should achieve
this optimal running time.

To expect a good self-improving algorithm that can  handle \emph{all}
distributions $\D$ seems a bit ambitious, and indeed we show that
even for the sorting problem there can be no space-efficient such 
algorithm (even when the entropy is low).  Hence, it seems
necessary to impose some kind of restriction on $\D$. However,
if we required $\D$ to be, say, uniform or a Gaussian, we would again
be stuck with the drawbacks of traditional average case analysis.
Hence, for self-improvement to be of any interest, the restricted
class of distributions should still be fairly general. One such class
is given by product distributions.

\subsection{Model and Results}

%The performance of a self-improving algorithm is measured
%with respect to an \emph{unknown} memoryless random 
%source $\D$  of input instances, where
%$\D = \prod_i \D_i$ is of \emph{product type}: each
%component $x_i$ of the input is drawn independently from an
%arbitrary, unknown distribution $\D_i$.
%The algorithm is given instances $I_0,I_1,\ldots$ drawn independently 
%from $\D$, which it must solve one at a time
%in batch mode with: (1) no prior knowledge of future instances, that 
%is, $f(I_k)$ 
%must be computed before any of the $I_j$'s ($j>k$) are known;
%and (2) no prior knowledge of $\D$.
%The algorithm may store auxiliary information
%to help improve its performance. (Unlike self-organizing
%data structures, however, none of that information
%should be \emph{necessary} for the algorithm to complete its task.)
%We use $\D$ as shorthand for ${\D}^n$,
%the $n$-th member of an infinite ensemble of distributions---one 
%for each input size. 
%After a training phase, we expect the algorithm
%to settle into its steady state whose expected running time
%is called its limiting complexity. Note that from the user's perspective
%the only difference noticeable in the training phase is that the system
%might be a little slower.

We will focus our attention on distributions $\D$ of 
\emph{product type}.
Think of each input as an $n$-dimensional vector $(x_1,\ldots,x_n)$ over
some appropriate domain. This could be a list of numbers (in the case of
sorting) or a list of points (for Delaunay triangulations). Each $x_i$
is generated independently at random from an arbitrary distribution $\D_i$,
so $\D = \prod_i \D_i$.
All the $\D_i$'s are independent of each other. It is fairly natural
to think of various portions of the input as being generated by independent
sources. For example, in computational geometry, the convex hull of uniformly 
independently distributed points in the unit square is a well studied problem.

Note that all our inputs are of the same size $n$. This might appear
to be a rather unnatural requirement for (say) a sorting algorithm.
Why must the 10th number in our input come from the same distribution?
We argue that this is not a major issue (for concreteness,
let us focus on sorting). The right way to think
of the input is as a set of sources $\D_1,\D_2,\ldots$,
each independently generating a single number. The actual ``order"
in which we get these numbers is not important. What \emph{is} important
is that for each number, we know its source. For a given input,
it is realistic to suppose that some sources may be active,
and some may not (so the input may have less than $n$ numbers).
Our self-improving sorters essentially perform an \emph{independent}
processing on each input number, after which $O(n)$ time is enough
to sort.\footnote{The self-improving Delaunay triangulation algorithms
have a similar behavior.} The algorithm is completely unaffected by the 
inactive sources. To complete the training phase, we only need to get
enough information about each source. 
What if new sources are introduced during the stationary
phase? Note that as long as $O(n/\log n)$ new sources (and hence
new numbers) are added, we can always include these extra numbers
in the sorted list in $O(n)$ time. Once the number of new sources
becomes too large, we will have to go back to the training phase. This
is, of course, quite acceptable: if the underlying distribution
of inputs changes significantly, we have to recalibrate
the algorithm. For these reasons, we feel that it is no loss
of generality to deal with a fixed input length, especially
for product distributions.

Our first result is a self-improving sorter.
Given a source $\D = \prod_i \D_i$ of real-number sequences 
$I = (x_1,\ldots,x_n)$,
let $\pi(I)$ denote the permutation induced by the ranks of the $x_i$'s,
using the indices $i$ to break ties. Observe that since $I$
is a random variable, so is $\pi(I)$.
We can define the entropy $H(\pi(I))$, over
the randomness of $\D$, and
the limiting complexity of our algorithm will depend on
$H(\pi(I))$. Note this quantity may be much smaller
than the entropy of the source itself but can never exceed it.

As we mentioned earlier, the self-improving algorithm
initially undergoes a training phase. At the end of this
phase, some data structures storing information about
the distributions are constructed. In the limiting phase,
the self-improving algorithm is fixed, and
these data structures do not change. In the context of sorting,
the self-improving sorter becomes some fixed comparison tree.
\medskip
\begin{theorem}\label{thm:sort-with-preprocessing}
There exists a self-improving sorter of 
$O(n+H(\pi(I)))$ limiting complexity,
for any input distribution $\D = \prod_i \D_i$.
Its worst case running time is $O(n\log n)$.
No comparison-based algorithm can sort an input from
$\D$ in less than $H(\pi(I))$ time.
For any constant $\eps>0$,
the storage can be made $O(n^{1+\eps})$ for 
an expected running time of $O(\eps^{-1}(n+ H(\pi(I))))$.
The training phase lasts $O(n^{\eps})$ rounds and the probability
that it fails is at most $1/n$.
\end{theorem}
\medskip

Why do we need a restriction on the input distribution?
In \S\ref{sec:lb}, we show that a self-improving sorter 
that can handle \emph{any} distribution requires an exponentially
large data structure. 
Fredman~\cite{Fredman76}
gave an algorithm that could optimally sort permutations from \emph{any} 
distribution $\D$. His algorithm needs to know $\D$ 
explicitly, and it constructs lookup tables of exponential size. 
Our bound shows that Fredman's algorithm cannot 
be improved.
Furthermore, we show that even for product distributions any
self-improving sorter needs super-linear space. Hence, our time-space
tradeoffs are essentially optimal. We remind the reader that we
focus on comparison-based algorithms.
\medskip
\begin{theorem} \label{thm:sort-lb} 
Consider a self-improving algorithm that, given any 
fixed distribution $\D$, can sort a random input from $\D$ in
expected $O(n + H(\pi(I)))$ time.
Such an algorithm requires $2^{\Omega(n\log n)}$ bits of storage.

Let $\eps \in (0,1)$.
Consider a self-improving algorithm that,
given any \emph{product distribution} $\D= \prod_i \D_i$,
can sort a random input from $\D$ in expected $\eps^{-1}(n + H(\pi(I)))$ time.
Such an algorithm requires a data structure 
of bit size $n^{1+\Omega(\eps)}$.
\end{theorem}
\medskip

For our second result, we take the notion of self-improving algorithms
to the geometric realm and address the classical
problem of computing the Delaunay triangulation
of a set of points in the Euclidean plane.
Given a source $\D = \prod_i \D_i$ of sequences 
$I = (x_1,\ldots,x_n)$ of points in $\R^2$,
let $T(I)$ denote the Delaunay triangulation of $I$.
If we interpret $T(I)$ as a random variable on the set of 
all undirected graphs with vertex set $\{1, \ldots, n\}$,
then $T(I)$ has an entropy $H(T(I))$, and
the limiting complexity of our algorithm depends on this entropy.
\medskip
\begin{theorem} \label{thm:del}
There exists a self-improving algorithm for planar Delaunay triangulations
of $O(n + H(T(I)))$ limiting complexity, for any input distribution 
$\D = \prod \D_i$. 
Its worst case running time is $O(n \log n)$.
For any constant $\eps > 0$, the storage can be made $O(n^{1+\eps})$ for
an expected running time of $O(\eps^{-1}(n + H(T(I))))$.
The training phase lasts $O(n^\eps)$ rounds and the probability
that it fails is at most $1/n$.
\end{theorem}
\medskip

From the linear time reduction from sorting to computing Delaunay 
triangulations~\cite[Theorems~8.2.2 and 12.1.1]{BoissonnatYv98}, 
the lower bounds of Theorem~\ref{thm:sort-lb} 
carry over to Delaunay triangulations. 

Both our algorithms follow the same basic strategy. 
During the training phase, we collect data about the inputs in 
order to obtain a \emph{typical}
input instance $V$ for $\D$ with $|V| = O(n)$, and we 
compute the desired structure $S$ (a sorted list or a Delaunay
triangulation) on $V$.
Then for each distribution $\D_i$, we construct an entropy
optimal search structure $D_i$ for $S$ (ie, an entropy optimal
binary search tree or a distribution sensitive planar point location 
structure).  In the
limiting phase, we use the $D_i$'s in order to locate the
components of a given input $I$ in $S$.
The fact that $V$ is a typical input ensures that $I$ will
be broken into individual subproblems of expected \emph{constant} size
that can be solved separately, so we can obtain the desired
structure for the input $V \cup I$ in expected linear time 
(plus the time for the $D_i$-searches). Finally, for 
both sorting and Delaunay triangulation it suffices to know
the solution for $V \cup I$ in order to derive the solution for
$I$ in linear expected time~\cite{CDH+,ChazelleMu09}. Thus, the running time 
of our algorithms
is dominated by the $D_i$-searches, and the heart of the analysis
lies in relating this search time to the entropies $H(\pi(I))$ and
$H(T(I))$, respectively. 

\subsection{Previous Work}

Related concepts to self-improving algorithms
have been studied before.
List accessing algorithms and splay trees
are textbook examples of how simple updating
rules can speed up searching with respect
to an adversarial request 
sequence~\cite{sleatorTpage,sleatorT85,
albersW98,borodinE,hesterH}.
It is interesting to note that 
self-organizing data structures were
investigated over stochastic input 
models first~\cite{albersM98,allenM78,bitner,gonnetMS,mccabe,rivest}.
It was the observation~\cite{bentleyM} that 
memoryless sources for list accessing are not terribly
realistic that partly motivated work on the adversarial models.
It is highly plausible that both approaches are superseded by
more sophisticated stochastic models: for example,
hidden Markov models for gene finding or speech recognition
or time-coherent models for self-customized BSP trees~\cite{arCT}
or for randomized incremental constructions~\cite{ChazelleMu09b}.
Recently, Afshani~\etal~\cite{AfshaniBaCh09} introduced
the notion of \emph{instance optimality}, which can be
seen as a generalization of output-sensitivity. They consider
the inputs as sets and try to exploit the structure \emph{within}
each input for faster algorithms.

Much research has been done on adaptive sorting~\cite{EstivillW},
especially on algorithms that exploit near-sortedness. 
Our approach is conceptually different: we seek to exploit
properties, not of individual inputs, but of their distribution.
Algorithmic self-improvement differs from past work on
self-organizing data structures and online computation in two 
fundamental ways.
First, there is no notion of an adversary: the inputs are generated by
a fixed, \emph{oblivious}, random source $\D$, and we compare ourselves
against an optimal comparison-based algorithm for $\D$. In 
particular, there is  no concept of competitiveness.
Second, self-improving algorithms do not exploit structure within 
any given input
but, rather, within the ensemble of input distributions. 

A simple example highlights this difference between previous sorters
and the self-improving versions. For $1 \leq i \leq n$, fix two random 
integers $a_i,b_i$
from $\{1,\ldots, n^2\}$. The distribution $\D_i$ is such
that $\Pr[x_i = a_i] = \Pr[x_i = b_i] = 1/2$, and
we take $\D = \prod_{i=1}^n \D_i$.
Observe that \emph{every} permutation generated by $\D$
is a random permutation, since the $a_i$'s and $b_i$'s are
chosen randomly.
Hence, any solution in the adaptive, self-organizing/adjusting framework
requires $\Omega(n\log n)$ time, because no input
$I_j$ exhibits any special structure to be exploited. 
On the other hand, our self-improving sorter
will sort a permutation from $\D$ in expected \emph{linear} time
during the limiting phase: since $\D$ generates
at most $2^n$ different permutations, we have $H(\pi(I)) = O(n)$.

\section{Entropy and Comparison-based Algorithms}\label{sec:entropy}

Before we consider sorting and Delaunay triangulations, let us first recall 
some 
useful properties of information theoretic entropy~\cite{CoverTh06} and 
explain how it relates to our notion of comparison-based algorithms.

Let $X$ be a random variable with a finite range $\cX$. The 
\emph{entropy} of $X$, $H(X)$, is defined as 
$H(X) \eqdef \sum_{x \in \cX} \Pr[X = x] \log(1/\Pr[X = x])$.
Intuitively, the event that $X = x$ gives us 
$\log(1/\Pr[X = x])$ bits of information about the underlying
elementary event, and $H(X)$ represents
the expected amount of information that can be obtained from observing
$X$. We recall the following well-known property of
the entropy of the Cartesian product
of independent random variables~\cite[Theorem~2.5.1]{CoverTh06}.
\medskip
\begin{claim}\label{clm:joint} 
Let $H(X_1,  \ldots, X_n)$
be the joint entropy of independent random variables $X_1, \ldots, X_n$.
Then  
\[ 
H(X_1,  \ldots, X_n) = \sum_i H(X_i).\qquad\endproof
\]
\end{claim}

We now define our notion of a comparison-based algorithm.
Let $\cU$ be an arbitrary universe, and let $\cX$ be a finite set.
A \emph{comparison-based algorithm} to compute a function 
$X : \cU \rightarrow \cX$ is a rooted binary tree $\cA$
such that (i) every internal node of $\cA$ represents a comparison of the 
form $f(I) \leq g(I)$, where $f,g : \cU \rightarrow \R$ are 
\emph{arbitrary} functions on the input universe $\cU$;
and (ii) the leaves of $\cA$ are labeled with outputs from $\cX$ such
that for every input $I \in \cU$, following the appropriate path for
$I$ leads to the correct output $X(I)$. If $\cA$ has maximum depth $d$, we
say that $\cA$ needs $d$ comparisons (in the worst case). For a
distribution $\D$ on $\cU$, the \emph{expected number of comparisons}
(with respect to $\D$) is the expected length of a path from
the root to a leaf in $\cA$, where the leaves are sampled according
to the distribution that $\D$ induces on $\cX$ via $X$.

Note that our 
comparison-based algorithms generalize both the traditional
notion of comparison-based algorithms~\cite[Chapter~8.1]{CormenLeRiSt09}, 
where the functions $f$ and $g$ are required to be 
projections, as well as the notion
of algebraic computation trees~\cite[Chapter~16.2]{AroraBa09}. Here the
functions $f$ and $g$ must be composed of elementary functions
(addition, multiplication, square root) such that the complexity  of 
the composition is proportional to the depth of the node. Naturally,
our comparison-based algorithms can be much stronger. For example,
deciding whether a sequence $x_1, x_2, \ldots, x_n$ of real numbers
consists of $n$ distinct elements needs \emph{one} comparison in our model,
whereas every algebraic computation tree for the problem
has depth $\Omega(n \log n)$~\cite[Chapter~16.2]{AroraBa09}. However,
for our problems of interest, we can still derive meaningful lower bounds.

\begin{claim}\label{clm:entropy-lower}
Let $\D$ be a distribution on a universe $\cU$ and let 
$X : \cU \rightarrow \cX$ be a random variable. Then any
comparison-based algorithm to compute $X$ needs 
at least $H(X)$ expected comparisons.
\end{claim}
\begin{proof}
This is an immediate consequence of Shannon's noiseless coding 
theorem~\cite[Theorem~5.4.1]{CoverTh06}
which states that any binary encoding
of an information source such as $X(I)$ must have an expected
code length of at least $H(X)$. Any comparison-based algorithm $\cA$
represents a coding scheme: the encoder sends the sequence of comparison
outcomes, and the decoder descends along the tree $\cA$, using the transmitted
sequence to determine comparison outcomes. Thus, any comparison-based
algorithm must perform at least $H(X)$ comparisons in expectation. 
\end{proof}
\medskip

Note that our comparison-based algorithms include all the 
traditional sorting algorithms~\cite{CormenLeRiSt09} (selection sort, 
insertion sort, quicksort, etc) as well as classic algorithms
for Delaunay triangulations~\cite{deBergKrOvSc00}
(randomized incremental construction, divide and conquer, plane sweep).
A notable exception are sorting algorithms that rely on table lookup
or the special structure of the input values (such as bucket sort or radix 
sort) as well as \emph{transdichotomous} algorithms for 
sorting~\cite{Han04,HanTh02} or Delaunay 
triangulations~\cite{BuchinMu09,ChanPa07,ChanPa09}.

The following lemma
shows how we can use the running times of comparison-based algorithms
to relate the entropy of different random variables. 
This is a very important tool that will be used to 
prove the optimality of our algorithms.
\medskip
\begin{lemma}\label{lem:relate-entropy}
Let $\D$ be a distribution on a universe $\cU$, and let
$X: \cU \rightarrow \cX$ and 
$Y: \cU \rightarrow \cY$
be two random variables. Suppose that the function 
$f$ defined by $f : (I, X(I)) \mapsto Y(I)$
can be computed by a comparison-based algorithm
with $C$ expected comparisons (where the expectation is over $\D$). 
Then $H(Y) = C + O(H(X))$, where all the entropies 
are with respect to $\D$.
\end{lemma}

\begin{proof}
Let $s : X(\cU) \rightarrow \{0,1\}^*$ be a unique binary encoding of 
$X(\cU)$. By unique encoding, we mean that the encoding is $1-1$. We denote the \emph{expected code length} of $s$ with
respect to $\D$, $\EX_\D[|s(X(I))|] $, by $E_s$.
By another application of Shannon's noiseless coding 
theorem~\cite[Theorem 5.4.1]{CoverTh06}), 
we have $E_s \geq H(X)$ for any unique encoding $s$ of $X(\cU)$, and there 
exists a unique encoding $s^*$ of $X(\cU)$ with $E_{s^*} = O(H(X))$. 

Using $f$, 
we can  convert $s^*$ into a unique encoding $t$ of $Y(\cU)$. Indeed, 
for every $I \in \cU$, $Y(I)$ can be uniquely identified by a string
$t(I)$ that is the concatenation of
$s^*(X(I))$ and additional bits that represent the 
outcomes of the comparisons for the algorithm to compute $f(I, X(I))$. 
Thus, for every element $y \in Y(\cU)$, we can define $t(y)$
as the lexicographically smallest string $t(I)$ for which $Y(I) = y$, and
we obtain a unique encoding $t$ for $Y(\cU)$. For the  expected code length
$E_t$ of $t$, we get
\[
E_t = E_\D[|t(Y(I))|] \leq 
E_\D[C + |s^*(X(I))| ] = C + E_{s^*} = C + O(H(X)). 
\]
Since Shannon's theorem implies $E_t \geq H(Y)$, 
the claim follows.
\end{proof}

\section{A Self-Improving Sorter}\label{sec:sorter}

We are now ready to describe our self-improving sorter. 
The algorithm takes an 
input $I = (x_1, x_2, \ldots, x_n)$ of real numbers drawn from
a distribution $\D = \prod_i \D_i$ (ie, each $x_i$ is chosen
independently from $\D_i$).
Let $\pi(I)$ denote the permutation induced by the ranks of the $x_i$'s,
using the indices $i$ to break ties.
By applying Claim~\ref{clm:entropy-lower} with $\cU = \R^n$, $\cX$
the set of all permutations on $\{1, \ldots, n\}$, and
$X(I) = \pi(I)$,  we see 
that any sorter must make at least $H(\pi(I))$ 
expected comparisons. Since it takes $\Omega(n)$ steps to
write the output, any sorter needs $\Omega(H(\pi(I)) + n)$ steps.
This is, indeed, the bound that our self-improving sorter achieves.

For simplicity, we begin with the steady-state algorithm
and discuss the training phase later.
We also assume that the distribution $\D$
is known ahead of time and that we are allowed some amount of preprocessing
before having to deal with the first input instance 
(\S\ref{sec:preprocessing-sort}).
Both assumptions are unrealistic, so we show how to remove them
to produce a bona fide self-improving sorter (\S \ref{sec:SI-sort}).
The surprise is how strikingly little of the distribution needs to be learned
for effective self-improvement.

\subsection{Sorting with Full Knowledge}\label{sec:preprocessing-sort}

We consider the problem of sorting $I= (x_1,\ldots,x_n)$, where each $x_i$ is
a real number drawn from a distribution $\D_i$. 
We can assume without loss of generality that all the $x_i$'s
are distinct. (If not, simply replace $x_i$ by $x_i +i\delta$ for 
an infinitesimally small $\delta > 0$, so that ties are broken according
to the index $i$.)

The first step of the self-improving sorter is to sample $\D$ a few times
(the training phase) and create a ``typical" instance to divide the real line
into a set of disjoint, sorted intervals.  Next, given some input $I$, the
algorithm sorts $I$ by using the typical instance, placing each input number in
its respective interval. All numbers falling into the same intervals are then
sorted in a standard fashion. The algorithm needs a few supporting data
structures.

\begin{itemize}
\item
\textsc{The $V$-list:}
Fix an integer parameter $\lambda = \lceil\log n\rceil$,
and sample $\lambda$ input instances from $\prod \D_i$. 
Form their union and sort the resulting $\lambda n$-element multiset
into a single list $u_1\leq \cdots \leq u_{\lambda  n}$.
Next, extract from it every $\lambda $-th item and form the list 
$V=(v_0,\ldots, v_{n+1})$, where $v_0=0$, $v_{n+1}=\infty$,
and $v_i= u_{i \lambda }$ for $1\leq i \leq n$.
Keep the resulting \emph{$V$-list} in a sorted table as 
a snapshot of a ``typical'' input instance.
We will prove the remarkable fact that,
with high probability, locating each $x_i$ in the $V$-list
is linearly equivalent to sorting $I$.
We cannot afford to search the $V$-list
directly, however. To do that, we need auxiliary search structures.

\item
\textsc{The $D_i$-trees:}
For any $i \geq 1$, let $\pred^V_i$ 
be the predecessor\footnote{Throughout this paper,
the predecessor of $y$ in a list refers to the index of
the largest list element $\leq y$; it does not
refer to the element itself.}
of a random $y$ from $\D_i$ 
in the $V$-list, and let $H_i^V$ be the entropy of 
$\pred^V_i$.
The \emph{$D_i$-tree} is defined to be an optimum binary search 
tree~\cite{Mehlhorn} over 
the keys of the $V$-list, where the access probability of 
$v_k$ is $\Pr_{\D_i}\bigl[x_i \in [v_k, v_{k+1})\bigr] = 
\Pr\bigl[\pred_i^V = k\bigr]$,
for any $0\leq k\leq n$.
This allows us to compute $\pred^V_i$
using $O(H_i^V +1)$ expected comparisons.

\end{itemize}

\begin{figure}
\begin{center}
\includegraphics{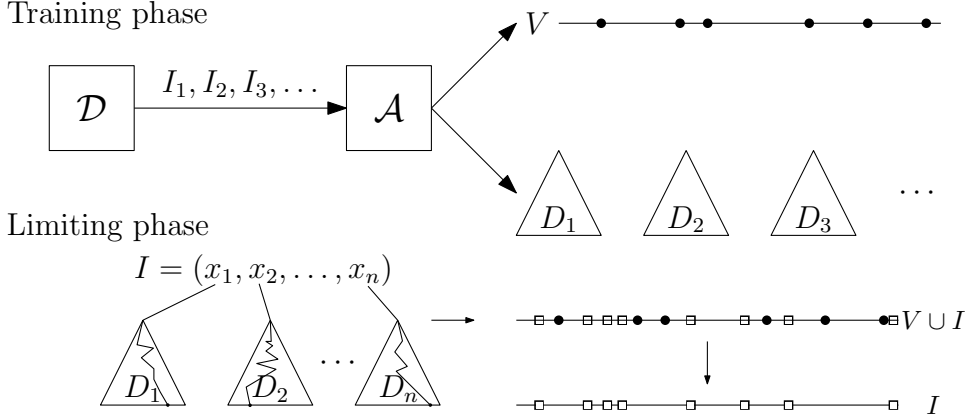}
\end{center}
\caption{The self-improving sorter: during the training phase, the algorithm
constructs a typical sorted list, the $V$-list, 
and a sequence $D_1$, $D_2$, $\ldots$
of optimal search trees for $V$ with respect to $\D_1$, $\D_2$, $\ldots$.
In the limiting phase, the algorithm uses the $D_i$'s to locate the 
$x_i$'s in the $V$-list, sorts the individual buckets, and removes the
elements from $V$.}
\label{fig:sisort}
\end{figure}

\textbf{The self-improving sorter.} The input $I$ is sorted 
by a two-phase procedure.  First we locate each $x_i$ in 
the $V$-list using the $D_i$-trees. This allows us to 
partition $I$ into groups $Z_0< Z_1 < \cdots$
of $x_i$'s sharing the same predecessor in the $V$-list.
The first phase of the algorithm takes $O(n+ \sum_i H_i^V)$ 
expected time.\footnote{The $H_i^V$'s themselves are 
random variables depending on the choice of the $V$-list. 
Therefore, this is a conditional expectation.}
The next phase involves going through each $Z_k$ and
sorting their elements naively, say using insertion sort, in
total time $O(\sum_k |Z_k|^2)$. See Fig.~\ref{fig:sisort}.
\medskip

The expected running time is $O(n+ \EX_\D[\sum_i H_i^V + \sum_k |Z_k|^2])$, and 
the total space used is $O(n^2)$. This can be decreased to $O(n^{1+\eps})$
for any constant $\eps > 0$; we describe how at the end of this section.
First, we show how to bound the running time of the first phase.
This is where we really show the optimality of our sorter.
\medskip
\begin{lemma} \label{lem:fredman-revisited}
\[ 
\sum_i H_i^V = O(n + H(\pi(I))). 
\]
\end{lemma}

\begin{proof}
Our proof actually applies to  \emph{any}
linear sized sorted list $V$. 
Let $\pred^V \eqdef (\pred_1^V, \ldots, \pred_n^V)$ be the
sequence of predecessors for all elements in $I$.
By Claim~\ref{clm:joint}, we have $H(\pred^V) = \sum_i H_i^V$,
so it suffices to bound the entropy of $H(\pred^V)$.
By Lemma~\ref{lem:relate-entropy} applied with
$\cU = \R^n$,   $X(I) = \pi(I)$ and $Y(I) = \pred^V$,
it suffices to give a comparison-based algorithm that can
determine $\pred^V(I)$ from $(I, \pi(I))$ with $O(n)$
comparisons. But this is easy: just use $\pi(I)$ to
sort $I$ (which needs no further comparisons) and then
merge the sorted list $I$ with $V$. Now the lemma follows
from Claim~\ref{clm:joint}.
\end{proof}
\medskip

Next we deal with the running time of the second phase. As long
as the groups $Z_k$ are small, the time to sort each group will
be small. The properties of the $V$-list ensure that this is the case.
\medskip
\begin{lemma} \label{lem:quicksort} 
For $0 \leq k \leq n$, let $Z_k =  \{x_i \mid v_k\leq x_i < v_{k+1}\}$
be the elements with predecessor $k$.
With probability at least $1 - n^{-2}$
over the construction of the $V$-list, we have 
$\EX_{\D}\big[|Z_k|\big]= O(1)$
and 
$\EX_{\D}\big[|Z_k|^2\big]= O(1)$, 
for all $0\leq k\leq n$.
\end{lemma}

\begin{proof} 
Remember that the $V$-list was formed by taking certain elements from a 
sequence $\hat I = s_1, s_2, \ldots,  s_{\lambda n}$ that was
obtained by concatenating $\lambda = \lceil \log n \rceil$ inputs 
$I_1$, $I_2$, $\ldots$.
Let $s_i \leq s_j$ be any two elements from $\hat I$, and
let $t = [s_i,s_j)$.
Note that all the other $\lambda n - 2$ numbers are
independent of $s_i$ and $s_j$.  Suppose we fix the values of $s_i$ and $s_j$
(in other words, we condition on the values of $s_i$ and $s_j$). For every 
$\ell \in \{1, \ldots, \lambda n\} \setminus \{i,j\}$, let
$Y^{(t)}_\ell$ be the indicator random variable for the event that 
$s_\ell \in t$, and let $Y^{(t)} \eqdef \sum_\ell Y^{(t)}_\ell$.  
Since all the
$Y^{(t)}_\ell$'s are independent, by Chernoff's 
bound~\cite[Theorem~4.2]{MotwaniRa95}, for any
$\beta \in [0,1]$,  
\begin{equation}\label{equ:chernoff}
\Pr[Y^{(t)} \leq (1-\beta)\EX[Y^{(t)}]] \leq 
\exp\bigl(-\beta^2\EX[Y^{(t)}]/2\bigr). 
\end{equation}
Setting $\beta = 10/11$, we see that
 if $\EX[Y^{(t)}] > 11\lceil\log n\rceil$,
then $Y^{(t)} > \lceil\log n\rceil$ with probability at least
$1 - 1/(\lambda^2n^4)$.
Note that this is true for \emph{every} fixing of $s_i$ and $s_j$.
Therefore, we get the above statement even with the unconditioned
random variable $Y^{(t)}$.
Now, by applying  the same argument
to any pair $s_i, s_j$ with $i \neq j$ and taking a union bound over all
$\binom{\lambda n}{2}$ such pairs, we get that with probability at 
least $1 - n^{-2}$ over the construction of $\hat I$
the following holds for all half-open intervals $t$ defined by pairs 
$s_i, s_j$ with $i \not= j$:
if $\EX[Y^{(t)}] > 11\lceil\log n\rceil$, then
$Y^{(t)} > \lceil \log n \rceil$. From now on we assume that this 
implication holds.

The $V$-list is constructed such that for $t_k = [v_k, v_{k+1})$, $Y^{(t_k)}
\leq \lceil\log n\rceil$, and hence 
$\EX[Y^{(t_k)}] = O(\log n)$.  Let $X^{(t_k)}_i$ be the indicator random 
variable 
for the event that $x_i \in_R \D_i$ lies in $t_k$, and $X^{(t_k)} \eqdef \sum_i
X^{(t_k)}_i = |Z_k|$.  Note that (where $a$ and $b$ denote the indices of 
$v_k$ and $v_{k+1}$ in $\hat I$)
\[
\EX[Y^{(t_k)}]  = 
\sum_{\ell \neq a, b}
\EX[Y^{(t_k)}_\ell]  \geq
\sum_{i} \lambda \EX[X^{(t_k)}_i] - 2
= \lceil \log n\rceil  \EX[X^{(t_k)}] -2,
\]
and therefore $\EX[X^{(t_k)}] =
O(1)$. Now, since the expectation of $X^{(t_k)}$ is constant, and
since $X^{(t_k)}$ is a sum of independent indicator random variables,
we can apply the following standard claim in order to show that the
second moment of $X^{(t_k)}$ is also constant.
\medskip
\begin{claim}\label{clm:indicator-square}
Let $X = \sum_i X_i$ be a sum of independent positive random variables with
$X_i = O(1)$ for all $i$ and $\EX[X] = O(1)$. Then $\EX[X^2] = O(1)$.
\end{claim}

\emph{Proof.}
By linearity of expectation,
\begin{multline*}
\EX\left[X^2\right]  = \EX\Bigl[\bigl(\sum_i X_i\bigr)^2\Bigr] = 
\sum_i \EX\left[X_i^2\right] + 2\sum_{i < j} \EX[X_i] \EX[X_j] \\
 \leq  \sum_i O\left(\EX[{X_i}]\right) + \Bigl(\sum_i \EX[X_i]\Bigr)^2 = O(1).
 \qquad \endproof
\end{multline*}
This concludes the proof of Lemma~\ref{lem:quicksort}.
\end{proof}
\medskip

Combining Lemmas~\ref{lem:fredman-revisited} and \ref{lem:quicksort}, 
we get the running time
of our self-improving sorter to be $O(n + H(\pi(I)))$.
This proves the optimality of time taken by the sorter.

We now show that the
storage can be reduced to $O(n^{1+\eps})$,
for any constant $\eps > 0$. The main idea is to prune each 
$D_i$-tree to depth $\eps\log n$. This ensures that 
tree has size $O(n^\eps)$, so the total storage used is
$O(n^{1+\eps})$. We also construct a completely balanced binary
tree $T$ for searching in the $V$-list. Now, when we wish to search
for $x_i$ in the $V$-list, we first search using the pruned $D_i$-tree.
At the end, if we reach a leaf of the \emph{unpruned} $D_i$-tree,
we stop since we have found the right interval of the $V$-list
which contains $x_i$. On the other hand, if the search in the $D_i$-tree
was unsuccessful, then we use $T$ for searching.

In the first case, the time taken for searching is
simply the same as with
unpruned $D_i$-trees. In the second case, the time
taken is $O((1+\eps)\log n)$. But note that the time
taken with unpruned $D_i$-trees is at least $\eps\log n$
(since the search on the pruned $D_i$-tree failed, we
must have reached some internal node of the unpruned tree).
Therefore, the extra time taken is only a $O(\eps^{-1})$ factor
of the original time. As a result, the space can be
reduced to $O(n^{1+\eps})$ with only a constant factor
increase in running time (for any fixed $\eps > 0$).

\subsection{Learning the Distribution}\label{sec:SI-sort}

In the last section we showed how to obtain a self-improving
sorter if $\D$ is known. We now explain how to remove this assumption.
The $V$-list is built in the first $\lceil \log n \rceil$ rounds, as
before.
The $D_i$-trees will be built after $O(n^\eps)$ additional rounds,
which will complete the training phase. During that phase,
sorting is handled via, say, mergesort to guarantee $O(n\log n)$ complexity.
The training part per se consists of learning basic information
about $\pred_i^V$ for each $i$. For notational simplicity, fix $i$
and let $p_{k}=  \Pr[\pred_i^V = k] = 
\Pr_{\, \D_i}\,[\, v_k \leq x_i < v_{k+1}\,]$.
Let $M = c n^\eps$, for a large enough constant $c$.
For any $k$, let $\chi_k$ be the number of times, 
over the first $M$ rounds, that $v_k$ is
found to be the $V$-list predecessor of %some 
$x_i$. (We use
standard binary search to compute predecessors in the training phase.)
Finally, define the $D_i$-tree to be a
weighted binary search tree defined over all the $v_k$'s such that
$\chi_k>0$. Recall that the crucial property of such
a tree is that the node associated with a key of weight $\chi_k$
is at depth $O(\log (M/\chi_k))$. 
We apply this procedure for each $i=1,\ldots, n$.

This $D_i$-tree is essentially the pruned version
of the tree we used in \S \ref{sec:algorithm}. Like before, its size is
$O(M)= O(n^\eps)$, and it is used in a similar
way as in \S \ref{sec:algorithm}, with a few minor differences. For 
completeness, we go over it again: given $x_i$, we perform a %binary
search down the $D_i$-tree. If we encounter a node
whose associated key $v_k$ is such that $x_i\in [v_k,v_{k+1})$,
we have determined $\pred_i^V$  and we stop the search.
If we reach a leaf of the $D_i$-tree without success, we
simply perform a standard binary search in the $V$-list.
\medskip
\begin{lemma} \label{lem:learn} 
Fix $i$. With probability at least $1-1/n^3$, for any $k$,
if $p_k > n^{-\eps/3}$ then $Mp_k/2 < \chi_k < 3Mp_k/2$.
\end{lemma}

\begin{proof}
The expected value of $\chi_k$ is $Mp_k$. If $p_k> n^{-\eps/3}$
then, by Chernoff's bound~\cite[Corollary~A.17]{AlonS} 
the count $\chi_k$ deviates from its expectation by more than 
$a = Mp_k/2$ with probability less than (recall that $M = cn^{\eps}$)
\[
2\exp(-2a^2/M) = 2\exp(-Mp_k^2/2) < 
2\exp(-(c/2)n^{2\eps/3}) \leq n^{-4},
\]
for $c$ large enough.
A union bound over all $k$ completes the proof.
%\qed
\end{proof}
\medskip

Suppose now the implication of Lemma~\ref{lem:learn} holds for all $k$ 
(and fixed $i$).
We show now that the expected search time for $x_i$ is 
$O(\eps^{-1} \,H_i^V +1)$.
Consider each element in the sum $H_i^V = \sum_k p_k\log (1/p_k)$.
We distinguish two cases.

\begin{itemize}
\item \textbf{Case 1: $p_k > n^{-\eps/3}$.} 
In this case, $v_k$ must be in $D_i$,
as otherwise we would have $\chi_k = 0$
by the definition of $D_i$, a contradiction (for $n$ large enough)
to Lemma~\ref{lem:learn},
which states that $\chi_k > Mn^{-\eps/3}/2$ .
Hence,
the cost of the search is $O(\log (M/\chi_k))$, and its
contribution to the expected search time is 
$O(p_k \log (M/\chi_k))$. By Lemma~\ref{lem:learn},
this is also $O(p_k(1+\log p_k^{-1}))$, as desired.

\item
\textbf{Case 2: $p_k \leq n^{-\eps}$.} The search time is always 
$O(\log n)$; hence the contribution to 
the expected search time is $O(\eps^{-1} p_k\log p_k^{-1})$.
\end{itemize}

By summing up over all $k$, we find that 
the expected search time is $O(\eps^{-1} \,H_i^V +1)$.
This assumes the implication of Lemma~\ref{lem:learn} for all $i$.
By a union bound, this holds with probability at least $1-1/n^2$.
The training phase fails when either this does not hold,
or if the $V$-list does not have the desired properties 
(Lemma~\ref{lem:quicksort}).
The total probability of this is at most $1/n$.
%
%, but this implication 
%holds for all $i$ with probability at least $1-1/n$.
%This leaves a probability $1/n$ that the training fails
%and we are stuck with $\Theta(n\log n)$ sorting---note 
%that we do not try to detect failure. However, this adds only an additive
%term of $O(\log n)$ to the expected complexity of our algorithm 
%and is therefore negligible.

\subsection{Lower Bounds} \label{sec:lb}

Can we hope for a result similar to Theorem~\ref{thm:sort-with-preprocessing}
if we drop the independence assumption? The short answer is no.
As we mentioned earlier, Fredman~\cite{Fredman76} gave a 
comparison-based algorithm that can optimally
sort any distribution of permutations. This uses 
an \emph{exponentially} large data
structure to decide which comparisons to perform. Our 
lower bound shows that the
storage used by Fredman's algorithm is essentially optimal.

To understand the lower bound, let us try to abstract out the behavior
of a self-improving sorter. Given inputs from a distribution $\D$, at each
round, the self-improving sorter is just a comparison tree for sorting.
After any round, the self-improving sorter may wish to update the 
comparison tree.
At some round (eventually), the self-improving sorter must
be able to sort with expected $O(n + H(\pi(I)))$ comparisons: 
the algorithm has ``converged" to the optimal comparison tree.
The algorithm uses some data structure to represent
(implicitly) this comparison tree.

We can think of a more general situation. The algorithm is explicitly
given an input distribution $\D$. It is allowed some space where it 
stores information about $\D$ (we do not care about the
time spent to do this). Then, (using this stored information)
it must be able to sort a permutation from $\D$ in 
expected $O(n + H(\pi(I)))$ comparisons.
So the information encodes some fixed comparison based procedure.
As a shorthand for the above, we will say that the \emph{algorithm,
on input distribution $\D$, optimally sorts $\D$}.
How much space is required to deal with all possible $\D$'s?
Or just to deal with product distributions? These are the
questions that we shall answer.
\medskip
\begin{lemma}\label{lem:exp-lower-bound} 
Let $h = (n\log n)/\alpha$, for some sufficiently large constant 
$\alpha < 0$, and let $\cA$ be an algorithm that can optimally sort any input 
distribution $\D$ with $H(\pi(I)) \leq h$.
Then $\cA$ requires $2^{\Omega(n \log n)}$ bits of storage.
\end{lemma}

\begin{proof}
Consider the set of all $n!$ permutations of $\{1, \ldots, n\}$.
Every subset $\Pi$ of $2^h$ permutations induces a distribution $\D^\Pi$
defined by picking every permutation in $\Pi$ with equal probability
and none other. Note that the total number such distributions $\D^\Pi$ is
$\binom{n!}{2^h} > (n!/2^h)^{2^h}$ and $H(\D_{<}^\Pi)= h$,
where $\D_{<}^\Pi$ is the distribution on the output $\pi(I)$ induced by
$\D^\Pi$.
Suppose there exists a comparison-based procedure ${\cA}_\Pi$ that sorts 
a random input from $\D^\Pi$ in expected time at most $c(n+h)$, 
for some constant $c>0$.
By Markov's inequality this implies that 
at least half of the permutations in $\Pi$ are sorted by 
${\cA}_\Pi$ in at most $2c(n+h)$ comparisons.
But, within $2c(n+h)$ comparisons, the procedure 
${\cA}_\Pi$ can only sort a set $P$ of at most $2^{2c(n+h)}$ permutations. 
Therefore, any other $\Pi'$ such that ${\cA}_{\Pi'} = {\cA}_\Pi$
will have to draw at least half of its elements from $P$.
This limits the number of such $\Pi'$ to 
\[ 
\binom{n!}{2^h/2}\binom{2^{2c(n+h)}}{2^h/2} < (n!)^{2^{h-1}}2^{c(n+h)2^h}. 
\]
This means that the number of distinct procedures needed
exceeds 
\[
(n!/2^h)^{2^h}/((n!)^{2^{h-1}}2^{c(n+h)2^h}) >
(n!)^{2^{h-1}}  2^{-(c+1)(n+h){2^h}} = 2^{\Omega(2^h n\log n)},
\]
assuming that $h/(n\log n)$ is small enough.
A procedure is entirely specified by a string of bits; therefore 
at least one such procedure must require storage logarithmic in the previous 
bound.
\end{proof}
\medskip

We now show that a self-improving sorter dealing with product
distributions requires super-linear size.
In fact, the achieved tradeoff between the $O(n^{1+\eps})$ storage bound 
and an expected running time off the optimal by a factor of $O(1/\eps)$ 
is optimal.
\medskip
\begin{lemma}\label{lem:space-lb}
Let $c > 0$ be a large enough parameter, and
let $\cA$ be an algorithm that, given a product 
distribution $\D$, can sort a random permutation from $\D$
in expected time $c(n+H(\pi(I)))$. Then $\cA$ requires 
a data structure of bit size $n^{1+\Omega(1/c)}$.
\end{lemma}

\begin{proof}
The proof is a specialization of the argument used for proving
Lemma~\ref{lem:exp-lower-bound}. Let $h = (n\log n)/(3c)$ and
$\kappa= 2^{\lfloor h/n \rfloor}$.
We define $\D_i$ by choosing $\kappa$ distinct integers in $\{1,\ldots, n\}$
and making them equally likely to be picked as $x_i$.
(For convenience, we use the tie-breaking rule that maps
$x_i\mapsto nx_i +i-1$. This ensures that $\pi(I)$ is unique.) We then 
set $\D \eqdef \prod_i \D_i$. By Claim~\ref{clm:joint}, $\D$ has
entropy $n\cdot \lfloor h/n \rfloor = \Theta(h)$.
This leads to $\binom{n}{\kappa}^n > (n/\kappa)^{\kappa n}$ choices of 
distinct distributions $\D$.
Suppose that $\cA$ uses $s$ bits of storage
and can sort each such distribution in $c(n+h)$ expected comparisons.
Some fixing $\cS$ of the bits must be able to accommodate
this running time for a set $\cG$ of 
at least $(n/\kappa)^{\kappa n}2^{-s}$ distributions $\D$.
In other words, some comparison-based procedure
can deal with $(n/\kappa)^{\kappa n}2^{-s}$ distributions $\D$.
Any input instance that is sorted in at most $2c(h+n)$ time
by $\cS$ is called \emph{easy}: the set of easy
instances is denoted by $\cE$.

Because $\cS$ has to deal with many distributions, there must
be many instances that are easy for $\cS$. This gives
a lower bound for $|\cE|$. On the other hand, 
since easy instances are those that are sorted extremely quickly by $\cS$,
there cannot be too many of them. This gives an upper bound for $|\cE|$.
Combining these two bounds, we get a lower bound for $s$. We will
begin with the easier part: the upper bound for $|\cE|$. 
\begin{claim} \label{clm:E-upper} 
$|{\cE}| \leq  2^{2c(h+n)+2}$
\end{claim} 

\emph{Proof:} 
In the comparison-based algorithm represented by $\cS$, 
each instance $I \in \cE$ is associated with a leaf
of a binary decision tree of depth at most $2c(h+n)$, ie, with one
of at most $2^{2c(h+n)}$ leaves. This would give us an upper bound
on $s$ if each $I \in \cE$ was assigned a distinct leaf.
However, it may well be that two distinct inputs $I, I' \in \cE$ have
$\pi(I) = \pi(I')$ and lead to the same leaf. Nonetheless, we have
a \emph{collision} bound, saying that for any permutation $\pi$,
there are at most $4^n$ instances  $I \in \cE$ with $\pi(I) = \pi$.
This implies that 
$|{\cE}|\leq  4^n2^{2c(h+n)}$.

To prove the collision bound, first fix a permutation $\pi$. How many
instances can map to this permutation? We argue 
that knowing that $\pi(I) = \pi$ for 
an instance $I \in \cE$, we only need
$2n-1$ additional bits to encode $I$. This immediately
shows that there must be less than $4^n$ such instances $I$.
Write $I = (x_1,\ldots, x_n)$,
and let $I$ be sorted to give the vector $\overline{I} = (y_1,\ldots, y_n)$.
Represent the ground set of $I$ as an $n$-bit vector $\alpha$
($\alpha_i = 1$ if some $x_j = i$, else $\alpha_i = 0$).
For $i= 2,\ldots, n$, let $\beta_i = 1$ if $y_i = y_{i-1}$,
else $\beta_i = 0$. Now, given $\alpha$ and $\beta$, we can immediately
deduce the vector $\overline{I}$, and by applying $\pi^{-1}$ to 
$\overline{I}$, we get $I$.
This proves the collision bound.
\qquad \endproof
\smallskip
\begin{claim} \label{clm:E-lower} 
$|{\cE}| \geq n^n \kappa^{-2n} 2^{-2s/\kappa}$
\end{claim} 

\emph{Proof:} 
Each $\D_i$ is characterized by a vector 
$v_i=(a_{i,1},\ldots,a_{i,\kappa})$, so that 
$\D$ itself is specified by 
$v= (v_1,\ldots, v_n)\in \R^{n\kappa}$. (From now on,
we view $v$ both as a vector and a distribution of input instances.)
Define the $j$-th projection of $v$ as $v^j=(a_{1,j},\ldots, a_{n,j})$.
Even if $v\in {\cG}$, 
it could well be that none of the 
projections of $v$ are easy. However, if we consider
the projections obtained by permuting the coordinates
of each vector $v_i=(a_{i,1},\ldots,a_{i,\kappa})$ in all possible
ways we enumerate each input instance from $v$ 
the same number of times. 
Note that applying these permutations gives us different vectors which also 
represent $\D$.
Since the expected time to sort an input chosen from $\D\in {\cG}$
is at most $c(h+n)$,
by Markov's inequality, there exists a choice of permutations 
(one for each $1 \leq i \leq n$) for which at least half of 
the projections of the vector obtained by applying these permutations are easy. 

Let us count how many distributions have a vector representation with a choice
of permutations placing half its projections in $\cE$.
There are fewer than $|{\cE}|^{\kappa /2}$
choices of such instances and, for any such choice, 
each $v'_i=(a_{i,1},\ldots,a_{i,\kappa})$ has half its entries
already specified, so the remaining choices are fewer
than $n^{\kappa n/2}$. This gives an upper bound of 
$n^{\kappa n/2} |{\cE}|^{\kappa/2}$ on the number of such distributions.
This number cannot be smaller than 
$|{\cG}| \geq (n/\kappa)^{\kappa n}2^{-s}$; therefore
%%
%\begin{equation*}\label{E}
$|{\cE}| \geq n^n \kappa^{-2n} 2^{-2s/\kappa}$, as desired. \qquad \endproof
%\end{equation*} 
\medskip

It now just remains to put the bounds together.
\begin{align*}
& n^n \kappa^{-2n} 2^{-2s/\kappa} &&\leq&& 2^{2c(h+n)+2} \\
\Longrightarrow\quad&  n\log n - 2n\log \kappa - 2s/\kappa &&\leq&& 2ch+2cn+2 \\
\Longrightarrow\quad&  \kappa n(\log n - 2\log \kappa) - 
2c\kappa h - 2c\kappa n - 2\kappa &&\leq&& 2s.
\end{align*}
We have $\kappa = n^{\Theta(1/c)}$ and $h = (n\log n)/(3c)$. 
Since $c$ is sufficiently large,
we get $s = n^{1+\Omega(1/c)}$.
\end{proof}

\section{Delaunay Triangulations}\label{sec:delaunay}

We now consider self-improving algorithms for Delaunay triangulations.
The aim of this section is to prove Theorem~\ref{thm:del}.
Let $I = (x_1,\ldots,x_n)$ denote an input instance, where each
$x_i$ is a point in the plane, generated
by a point distribution $\D_i$. The distributions $\D_i$
are arbitrary, and may be continuous, although we never
explicitly use such a condition.
Each $x_i$ is independent of the others, so in each round
the input $I$ is drawn from the product distribution
$\D = \prod_i \D_i$,  
and we wish to compute the Delaunay
triangulation of $I$, $T(I)$. 
To keep our arguments simple,
we will assume that the points of $I$ are in \emph{general position} 
(ie, no four points in $I$ lie on a common circle).
This is no loss of generality and does not restrict the distribution
$\D$, because the general position assumption can always be enforced
by standard symbolic perturbation techniques~\cite{EdelsbrunnerMu90}.
Also we will assume that there is a bounding triangle that always contains 
all the points in $I$. Again, this does not
restrict the distribution $\D$ in any way, because we can always 
simulate the bounding triangle symbolically by adding virtual points
at infinity.

The distribution $\D$ induces
a (discrete) distribution on the set of Delaunay triangulations,
viewed as undirected graphs with vertex set $\{1,\ldots, n\}$.  
Consider the entropy
of this distribution: for each graph $G$
on $\{1,\ldots, n\}$, let $p_G$ be the probability that it represents
the Delaunay triangulation of $I \in_R \D$. We have 
the output entropy $H(T(I)) \eqdef - \sum_G p_G \log p_G$.
By Claim~\ref{clm:entropy-lower}, 
any comparison-based algorithm 
to compute the Delaunay triangulation of $I \in_R \D$ needs
at least $H(T(I))$ expected comparisons.
Hence, an \emph{optimal} algorithm will be one that has an
expected running time of $O(n + H(T(I)))$ (since it takes
$O(n)$ steps to write the output).

We begin by describing the basic self-improving algorithm.
(As before, we shall first assume that some aspects of the 
distribution $\D$ are known.) Then,
we shall analyze the running time using our information theory tools
to argue that the expected running time is optimal. 
Finally, we remove the assumption that $\D$ is known and
give the time-space tradeoff in Theorem~\ref{thm:del}.

\subsection{The algorithm}\label{sec:algorithm}

We describe the algorithm in two parts. The first part explains 
the learning phase and the data structures that are constructed 
(\S \ref{sec:learning}).
Then, we explain how these data structures are used to
speed up the computation in the limiting phase (\S \ref{sec:lim}). 
As before, the expected running
time will be expressed in terms of certain parameters of the
data structures obtained in the learning phase. In the next
section (\S \ref{sec:running-time}), we will prove that these parameters are 
comparable to the output 
entropy $H(T(I))$. First, we will assume 
that the distributions $\D_i$ are known to
us, and the data structures described will
use $O(n^2)$ space.
Section~\ref{sec:tradeoff} repeats the arguments
of \S \ref{sec:SI-sort} to remove this assumption and to
give the space-time tradeoff bounds of Theorem~\ref{thm:del}.

As outlined in Fig.~\ref{fig:analogies}, our algorithm
for Delaunay triangulation is roughly a generalization of
our algorithm for sorting.  This is not surprising, but
note that while the steps of the two algorithms, and their
analyses, are analogous, in several cases a step for sorting
is trivial, but the corresponding step for Delaunay triangulation
uses some relatively recent and sophisticated prior work.

\begin{figure}
\begin{center}
\begin{tabular}{|p{170pt}|p{170pt}|}
    
    \hline   Sorting
            & Delaunay Triangulation
    \\ \hline Intervals $(x_i, x_{i'})$ containing no values of $I$
            & Delaunay disks
    \\ \hline Typical set $V$
            &Range space $\eps$-net $V$ \cite{MSW,CV}, ranges are disks, 
	    $\eps = 1/n$
    \\ \hline $\log n$ training instance points with the same $\pred_V$ value
            & $\log n$ training instance points in each Delaunay disk
    \\ \hline Expect $O(1)$ values of  $I$ within each bucket 
    (of the same $\pred^V$ index)
            & Expect $O(1)$ points of $I$ in each Delaunay disk of $V$
    \\ \hline Optimal weighted binary trees $D_i$
            & Entropy-optimal planar point location data structures 
	    $D_i$ \cite{AMM}
    \\ \hline Sorting within buckets
            & Triangulation within $\cV(Z_s) \cap s$ 
	    (Claim \ref{clm:VDVX})
    \\ \hline Sorted list of $V \cup I$
            & $T(V \cup I)$
    \\ \hline Build sorted $V$ from sorted $V \cup I$ (trivial)
            & Build $T(I)$ from $T(V \cup I)$ \cite{CDH+,ChazelleMu09}
    \\ \hline (analysis) merge sorted $V$ and $I$
            & (analysis) merge $T(V)$ and $T(I)$ \cite{C}
    \\ \hline (analysis) recover the indices $\pred^V_i$ from the sorted $I$ 
    (trivial)
            & (analysis) recover the triangles $\tsearch^V_i$ in $T(V)$ 
	    from $T(I)$ (Lemma~\ref{lem:computeconflict})
    \\ \hline
\end{tabular}
\end{center}
\caption{Delaunay triangulation algorithm as a generalization of the 
sorting algorithm}
\label{fig:analogies}
\end{figure}

\subsubsection{Learning Phase}\label{sec:learning}

For each round in the learning phase, we use a standard algorithm
to compute the output Delaunay triangulation. We also
perform some extra computation to build some
data structures that will allow speedup in the limiting phase.

The learning phase is as follows. Take the first
$\lambda \eqdef \lceil\log n\rceil$ input lists
$I_1$, $I_2$, $\ldots$, $I_\lambda$. 
Merge them into one list $\hat I$ 
of $\lambda n =  n\lceil\log n\rceil$ points. Setting $\eps \eqdef 1/n$, 
find an $\eps$-net $V\subseteq \hat I$ for the set of all open disks. In 
other words, find a set $V$ such that for any open disk $C$ that
contains more than $\eps \lambda n = \lceil \log n\rceil$ points
of $\hat I$, $C$ contains at least one point of $V$. 
It is well known that that there exist $\eps$-nets of
size $O(1/\eps)$ for disks~\cite{CV,MSW,Matousek92,PyrgaRa08},
which here is $O(n)$.
Furthermore, it is folklore that 
our desired $\eps$-net $V$ can be constructed in time
$n (\log n)^{O(1)}$, but there seems to
be no explicit description of such an algorithm 
for our precise setting. Thus, we present 
an algorithm based on a construction by
Pyrga and Ray~\cite{PyrgaRa08} in Appendix~\ref{app:epsnet}

Having obtained $V$, we construct the Delaunay triangulation of $V$,
which we denote by $T(V)$. 
This is the analog of the $V$-list for
the self-improving sorter.
We also build an optimal
planar point location structure (called $D$) for $T(V)$:
given a point, we can find in $O(\log n)$ time the triangle of $T(V)$ that
it lies in~\cite[Chapter~6]{deBergKrOvSc00}. Define the random variable 
$\tsearch_i^V$ to be the triangle of $T(V)$ that $x_i$ falls 
into.\footnote{Assume that we add the vertices of the bounding triangle to $V$.
This will ensure that $x_i$ will always fall in some triangle $\tsearch_i^V$.}
Now let the entropy of $\tsearch_i^V$ be $H^V_i$.
If the probability that $x_i$ falls in triangle $t$ of $T(V)$ is $p^t_i$,
then $H^V_i = -\sum_t p^t_i \log p^t_i$. 
For each $i$, we construct a search structure $D_i$ of size $O(n)$
that finds $\tsearch_i^V$ in expected $O(H^V_i)$ time.
These $D_i$'s can be constructed using 
the results of Arya~\etal~\cite{AMM}, for which the expected number
of primitive comparisons is $H^V_i + o(H^V_i)$.
These correspond to the $D_i$-trees used
for sorting.

\begin{figure}
\begin{center}
\includegraphics{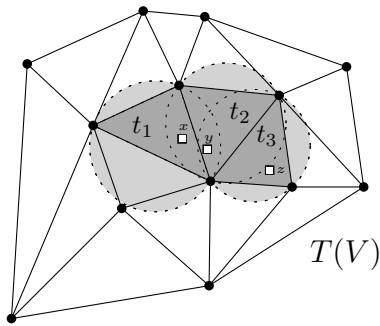}
\end{center}
\caption{Conflicts between $T(V)$ and the inputs: the input point $x$ conflicts with
triangles $t_1$ and $t_2$, $y$ conflicts with $t_1$, $t_2$, and $t_3$,
and $z$ conflicts only with $t_3$. } \label{fig:conf}
\end{figure}

We will now prove an analog to
Lemma~\ref{lem:quicksort} which shows that the
triangles of $T(V)$ do not contain many points
of a new input $I \in_R \D$ on the average. Consider
a triangle $t$ of $T(V)$ and let $C_t$ be its
circumscribed disk; $C_t$ is a Delaunay disk of $V$. If a point
$x_i \in I$ lies in $C_t$, we say that $x_i$ is \emph{in conflict}
with $t$ and call $t$ a \emph{conflict triangle} for $x_i$.
Refer to Fig.~\ref{fig:conf}.
(The ``conflict'' terminology arises from the fact that if
$x_i$ were added to $V$, triangles with which it conflicts
would no longer be in the Delaunay triangulation.) 
Let
$Z_t \eqdef I \cap C_t$, the random variable that represents
the points of $I \in_R \D$ that fall inside $C_t$, the
\emph{conflict set} of $t$. Furthermore, let
$X_t \eqdef |Z_t|$.
Note that the randomness comes from the random distribution
of $\hat I$ (on which $V$ and $T(V)$ depend), as well as the randomness of $I$.
We are interested in the expectation $\EX[X_t]$ over $I$ of $X_t$.
All expectations are taken over a random input $I$ chosen
from $\D$.

\begin{lemma} \label{lem:learning} 
For any triangle $t$ of $T(V)$,
let $Z_t = \{x_i \mid x_i \in C_t \}$ be the conflict set of $t$, 
and define $X_t \eqdef |Z_t|$.
With probability at least $1 - n^{-2}$ over
the construction of $T(V)$,
we have
$\EX[X_t] = O(1)$ and $\EX[X_t^2] = O(1)$, for all triangles $t$ of $T(V)$.
\end{lemma}

\begin{proof} This is similar to the argument
given in Lemma~\ref{lem:quicksort} with a geometric twist.
Let the list of points $\hat I$
be $s_1, \ldots, s_{\lambda n}$, the concatenation of $I_1$ through 
$I_\lambda$.
Fix three distinct indices $i,j,k$ and 
the triangle $t$ with vertices
$s_i, s_j, s_k$ (so we are effectively conditioning on $s_i, s_j, s_k$). Note
that all the remaining $\lambda n-3$ points are chosen independently
of $s_i, s_j$, $s_k$, from some distribution $\D_\ell$. For
each $\ell \in \{1,\ldots, \lambda n\} \setminus \{i,j,k\}$, let 
$Y^{(t)}_\ell$ be the indicator variable
for the event that $s_\ell$ is inside $C_t$. Let 
$Y^{(t)} = \sum_\ell Y^{(t)}_\ell$.
Setting $\beta = 11/12$ in (\ref{equ:chernoff}),
we get that
if $\EX[Y^{(t)}] > 12\lceil\log n\rceil$,
then $Y^{(t)} > \lceil\log n\rceil$
with probability at least $1 - 1/(\lambda^3 n^{5})$.
This is true for every fixing of $s_i,s_j,s_k$, so
it is also true unconditionally.
By applying the same argument to any triple $i,j,k$ of
distinct indices, and taking a union bound over all
$\binom{\lambda n}{3}$ triples,
we obtain that
with probability at least $1 - n^{-2}$, for any triangle
$t$ generated by the points of $\hat I$, if 
$\EX[Y^{(t)}] > 12\lceil\log n\rceil$,
then $Y^{(t)} > \lceil\log n\rceil$. We henceforth assume that this
event happens.

Consider a triangle $t$ of $T(V)$ and its circumcircle $C_t$.
Since $T(V)$ is Delaunay, $C_t$ contains no point of $V$
in its interior. Since $V$ is a $(1/n)$-net for all disks with
respect to $\hat I$, $C_t$ contains at most $\lceil\log n\rceil$ points
of $\hat I$, that is, $Y^{(t)}\le \lceil\log n\rceil$.
This implies that $\EX[Y^{(t)}] = O(\log n)$, as in the previous paragraph.
Since $\EX[Y^{(t)}] > \log n \EX[X_t]-3$, we obtain $\EX[X_t] = O(1)$,
as claimed. Furthermore, since $X_t$ can be written as a sum of independent 
indicator random variables, Claim~\ref{clm:indicator-square}
shows that $\EX[X_t^2] = O(1)$. 
\end{proof}

\subsubsection{Limiting Phase} \label{sec:lim}

We assume that we are done with the learning phase, and have $T(V)$
with the property given in Lemma~\ref{lem:learning}: for
every triangle $t \in T(V)$, $\EX[X_t] = O(1)$ and $\EX[X_t^2] = O(1)$. We have reached
the limiting phase where the algorithm is expected
to compute the Delaunay triangulation with the optimal running
time. We will prove the following lemma in this section.
\medskip

\begin{lemma} \label{lem:time} 
Using the data structures from
the learning phase, and the properties of them that hold
with probability at least $1-1/n^2$, in the limiting phase the 
Delaunay triangulation
of input $I$ can be generated in expected $O(n + \sum_{i=1}^n H^V_i)$ time.
\end{lemma}

The algorithm, and the proof of this lemma, has two steps.  In
the first step, $T(V)$ is used to quickly compute $T(V\cup I)$,
with the time bounds of the lemma.  In the second step,
$T(I)$ is computed from $T(V\cup I)$, using a
randomized splitting algorithm proposed by Chazelle~\etal~\cite{CDH+},
who provide the following theorem.
%\medskip
\begin{theorem} \cite[Theorem~3]{CDH+} 
Given a set of $n$ points $P$ and its Delaunay 
triangulation, for any partition of $P$ into two disjoint subsets $P_1$ 
and $P_2$, the Delaunay triangulations $T(P_1)$ and $T(P_2)$ can be computed 
in $O(n)$ expected time, using a randomized algorithm.
\end{theorem}

\begin{figure}
\begin{center}
\includegraphics{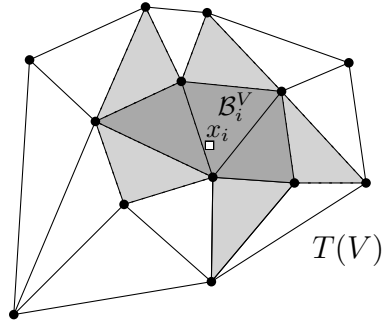}
\end{center}
\caption{Determining the conflict set for $x_i$:
the triangle $\tsearch_i^V$ containing $x_i$ is found via $D_i$.
Then we perform a breadth-first search from $\tsearch_i^V$ until
we encounter triangles that no longer conflict with $x_i$.
The dark gray triangles form the conflict set of $x_i$, the 
light gray triangles mark the end of the BFS. Since the conflict
set $S_i$ is connected, and since the dual graph has
bounded degree, this takes $O(|S_i|)$ steps.} \label{fig:bfs}
\end{figure}

The remainder of this section 
is devoted to showing that $T(V\cup I)$ can be computed in 
expected time $O(n + \sum_{i=1}^n H^V_i)$.
The algorithm is as follows.
For each $x_i \in I$, we use $D_i$ to find the 
triangle $\tsearch_i^V$ of $T(V)$ that
contains it. By the properties of the $D_i$'s as described in
\S\ref{sec:learning}, 
this takes $O(\sum_{i=1}^n H^V_i)$ expected time. We now need
to argue that given the $\tsearch_i^V$'s, the Delaunay triangulation
$T(V \cup I)$ can be computed in expected linear time.
For each $x_i$, we walk through
$T(V)$ and find all the Delaunay disks of $T(V)$ that
contain $x_i$, as in incremental
constructions of Delaunay triangulations~\cite[Chapter~9]{deBergKrOvSc00}.
This is done by breadth-first search of the 
dual graph of $T(V)$, starting from $\tsearch_i^V$.
Refer to Fig.~\ref{fig:bfs}.
Let $S_i$ denote the set of triangles whose circumcircles contain
$x_i$. We remind the reader that $Z_t$ is the conflict set
of triangle $t$.

\begin{claim} \label{clm:getting-si} 
Given all $\tsearch_i^V$'s, all $S_i$ and $Z_t$ sets 
can be found in expected linear time.
\end{claim}

\begin{proof} 
To find all Delaunay disks containing $x_i$, do a breadth-first search
from $\tsearch_i^V$. For any triangle $t$ encountered, check
if $C_t$ contains $x_i$. If it does not, then
we do not look at the neighbors of $t$. Otherwise, add 
$t$ to $S_i$ and $x_i$ to $Z_t$ and continue.
Since $S_i$ is connected in the dual graph of 
$T(V)$,\footnote{Since the triangles in $S_i$ cover exactly
the planar region of triangles incident to $x_i$ in
$T(V \cup \{x_i\})$.} we will visit
all $C_t$'s that contain $x_i$.
The time taken
to find $S_i$ is $O(|S_i|)$.
The total time taken
to find all $S_i$'s (once all the $\tsearch_i^V$'s are found)
is $O(\sum_{i=1}^n |S_i|)$. Define the indicator function $\chi(t,i)$ that
takes value $1$ if $x_i \in C_t$ and zero otherwise.  We have
\[
\sum_{i=1}^n |S_i| = \sum_{i=1}^n \sum_{t\in T(V)} \chi(t,i) = \sum_{t\in T(V)} \sum_{i=1}^n \chi(t,i) = \sum_t X_t.
\]
Therefore, by Lemma~\ref{lem:learning},
\[ 
\EX\Bigl[\sum_{i=1}^n |S_i|\Bigr] = 
\EX\Bigl[\sum_t X_t\Bigr] = \sum_t \EX[X_t] 
= O(n).
\]
This implies that all $S_i$'s and $Z_t$'s can
be found in expected linear time.
\end{proof}
\medskip

\begin{figure}
\begin{center}
\includegraphics{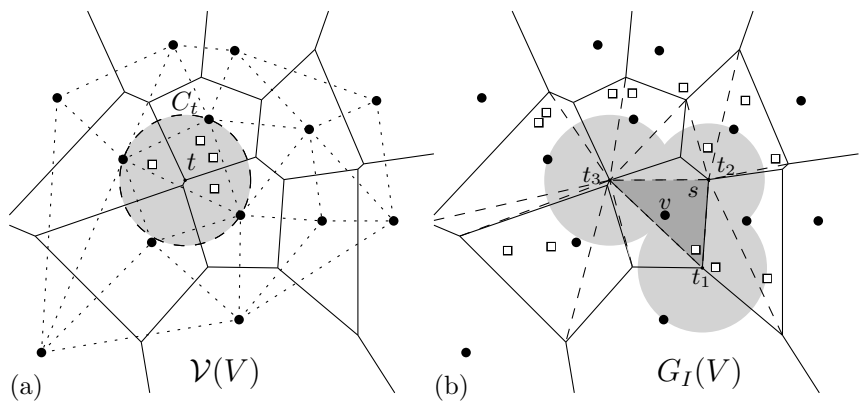}
\end{center}
\caption{(a) $\cV(V)$ is dual to $T(V)$. Each vertex $t$ of $\cV(V)$ 
corresponds to the center of the circumcircle of a triangle $t$ of $T(V)$,
and it has the same conflict set $Z_t$ of size $X_t$. 
(b) The geode triangulation
$G_I(V)$ is obtained by connecting the vertices of each region of $\cV(V)$ to
the lexicographically smallest incident vertex with the smallest $X_t$. 
The conflict set of
a triangle $s$ is the union of the conflict sets of its vertices and
point $v$ defining the region.} \label{fig:vor}
\end{figure}
Our aim is to build the Delaunay triangulation
$T(V \cup I)$ in linear time using the conflict sets $Z_t$. To that end,
we will use divide-and-conquer to compute the \emph{Voronoi diagram}
$\cV(V \cup I)$,
using a scheme that has been used for nearest neighbor searching \cite{C88}
and for randomized convex hull constructions~\cite{CS89,Chazelle00}.
It is well
known that the Voronoi diagram of a point set is dual to the Delaunay 
triangulation,
and that we can go from one to the other in linear 
time~\cite[Chapter~9]{deBergKrOvSc00}.
Refer to Fig.~\ref{fig:vor}(a).
Consider the Voronoi diagram of $V$, $\cV(V)$.
By duality, the vertices of $\cV(V)$ correspond to the triangles
in $T(V)$, and we identify the two. In particular, each vertex $t$ of 
$\cV(V)$ has a conflict set $Z_t$, the conflict set for the 
corresponding triangle in $T(V)$, and $|Z_t| = X_t$, by our definition 
of $X_t$ (see Fig.~\ref{fig:vor}(a)).
We triangulate the Voronoi diagram as follows: for each region $r$ of
$\cV(V)$, determine the lexicographically smallest Voronoi vertex 
$t_r$ in $r$ with minimum $X_t$. Add edges from  all the Voronoi vertices 
in $r$ to $t_r$. Since each region of $\cV(V)$ is convex, this 
yields a triangulation\footnote{We need to be a bit careful when handling 
unbounded 
Voronoi regions:
we pretend that there is a Voronoi vertex $p_\infty$ at infinity which is the 
endpoint of all unbounded Voronoi edges, and when we triangulate the 
unbounded region, we also add
edges to $p_\infty$. By our bounding triangle assumption, there is no point in 
$I$ outside the convex hull of $V$ and hence the conflict set 
of $p_\infty$ is empty.} of $\cV(V)$. We call it the 
\emph{geode triangulation} of $\cV(V)$ with respect to $I$, 
$G_I(V)$~\cite{C88,Chazelle00}. %(see Figure 2 of~\cite{C88}).
Refer to Fig.~\ref{fig:vor}(b).
Clearly, $G_I(V)$ can be computed in linear time. 
We extend the notion of conflict set to the triangles
in $G_I(V)$: Let $s$ be a triangle in $G_I(V)$ and let $t_1$, $t_2$, $t_3$ be
its incident Voronoi vertices. Then the conflict set of $s$, $Z_s$, 
is defined as 
$Z_s \eqdef Z_{t_1} \cup Z_{t_2} \cup Z_{t_3} \cup \{v\}$, 
where $v \in V$ is the
point whose Voronoi region contains the triangle $s$.
In the following, for any two points $x$ and $y$,
$|x-y|$ denotes the Euclidean distance between them.
\medskip
\begin{claim}\label{clm:VDVX}
Let $s$ be a triangle of $G_I(V)$ and let $Z_s$ be its conflict set. Then the
Voronoi diagram of $V \cup I$ restricted to $s$, 
$\cV(V \cup I) \cap s$,
is the same as the Voronoi diagram of $Z_s$ restricted to $s$, 
$\cV(Z_s) \cap s$.
\end{claim}

\begin{figure}
\begin{center}
\includegraphics{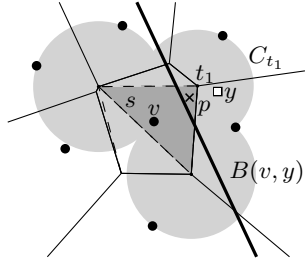}
\end{center}
\caption{The nearest neighbor of a point $y \in s$ is either
$v$ or needs to be in the conflict set of one of its vertices.} \label{fig:nbr}
\end{figure}

\begin{proof}
Consider a point $p$ in the triangle $s$, and let $y$ be
the nearest neighbor of $p$ in $V \cup I$. If $y \in V$, then
$y$ has to be $v$, since $s$ lies in the Voronoi region of $v$
with respect to $V$. Now suppose that $y \in I$. 
Let $B(v,y)$ be the \emph{perpendicular bisector} of the line 
segment $(v,y)$ (ie, the line
containing all points in the plane that have equal distance from
$v$ and $y$). Refer to Figure~\ref{fig:nbr}. Let $B^+$ be the halfplane 
defined by $B(v,y)$ that contains $y$. Since $B^+$ intersects $s$, by 
convexity it also contains a vertex of $s$, say $t_1$. Because $t_1$ 
and $y$ are on the same side ($B^+$),
$|y - t_1| < |v - t_1|$.
Note that $C_{t_1}$ has center $t_1$ and radius $|v-t_1|$, because
$t_1$ is a vertex of the Voronoi region corresponding
to $v$ (in $\cV(V)$).
Hence, $y \in Z_{t_1}$.
It follows that $y \in Z_s$, so 
$\cV(V \cup I) \cap s =  \cV(Z_s) \cap s$, as claimed.
\end{proof}
\medskip

Claim~\ref{clm:VDVX} implies that $\cV(V \cup I)$ can be found
as follows: for each triangle $s$ of $G_I(V)$, compute 
$\cV(Z_s) \cap s$, the Voronoi diagram of $Z_s$ restricted to $s$. 
Then, traverse the edges of $G_I(V)$ and fuse the bisectors of the
adjacent diagrams, yielding $\cV(V \cup I)$.
\begin{lemma}
Given $\cV(V)$, the Voronoi diagram $\cV(V \cup I)$ can be 
computed in expected $O(n)$ time.
\end{lemma}

\begin{proof}
The time to find $\cV(Z_s) \cap s$ for a triangle 
$s$  in $G_I(V)$ is 
$O(|Z_s|\log |Z_s|) = O(|Z_s|^2)$~\cite[Chapter~7]{deBergKrOvSc00}. 
For a region $r$ of 
$\cV(V)$, let $S(r)$ denote the set of triangles of $G_I(V)$ contained
in $r$, and let $E(r)$ denote the set of edges in $\cV(V)$ 
incident to $r$. 
Recall that $t_r$ denotes the common vertex of all triangles in
$S(r)$. The total running time is
$O(\EX\Bigl[\sum_{s \in G_I(V)} |Z_s|^2\Bigr])$, which is
proportional to
\begin{multline*}
%\EX\Bigl[\sum_{s \in G_I(V)} |Z_s|^2\Bigr] =
\EX\Bigl[\sum_{r \in \cV(V)} \sum_{s \in S(r)} |Z_s|^2\Bigr] 
\leq \EX\Bigl[\sum_{r \in \cV(V)} \sum_{
(t_1,t_2) \in E(r)}
(1+X_{t_r} + X_{t_1} + X_{t_2})^2\Bigr]\\
\leq \EX\Bigl[\sum_{r \in \cV(V)} 
\sum_{(t_1, t_2) \in E(r)} 
(1+2X_{t_1} + X_{t_2})^2\Bigr],
\end{multline*}
since $X_{t_r} \leq \min(X_{t_1}, X_{t_2})$. 
For $e = (t_1,t_2)$, let $Y_e = 1+2X_{t_1} + X_{t_2}$.
Note that $\EX[Y_e] = O(1)$, by Lemma~\ref{lem:learning}.
We can write $Y_e = \sum_{i} (1/n + 2\chi(t_1,i) + \chi(t_2,i))$,
where $\chi(t,i)$ was the indicator random variable for the event
that $x_i \in C_t$.
Hence, since $1/n + 2\chi(t_1,i) + \chi(t_2,i) < 4$,
Claim~\ref{clm:indicator-square} implies that $\EX[Y_e^2] = O(1)$.
Thus,
%\begin{align*}
\[
\EX\Bigl[\sum_{s \in G_I(V)} |Z_s|^2\Bigr]
\leq
\sum_{r \in \cV(V)} 
\sum_{\substack{e \in E(r) \\ e = (t_1, t_2)}} 
\EX[(Y_e)^2]
=\sum_{r \in \cV(V)} 
\sum_{\substack{e \in E(r) \\ e = (t_1, t_2)}} 
O(1).
\]
%\end{align*}

The number of edges in $\cV(V)$ is linear, and each edge $e$
is incident to exactly two Voronoi regions $r$.
Therefore, $\EX[\sum_{s \in G_I(V)} |Z_s|^2] = O(n)$.
Furthermore, assembling the restricted diagrams takes time 
$O\bigl(\EX\bigl[\sum_{s \in G_I(V)}|Z_s|\bigr]\bigr)$, and as
$|Z_s| \leq |Z_s|^2$, this is also linear.  
\end{proof}

\subsection{Running time analysis}\label{sec:running-time}

In this section, we prove that the running time bound
in Lemma~\ref{lem:time} is indeed optimal. As discussed 
at the beginning of \S\ref{sec:delaunay}, 
Claim~\ref{clm:entropy-lower} implies that any
comparison-based algorithm for computing the Delaunay
triangulation of input $I \in_R \D$ needs at least
$H(T(I))$ expected comparisons.
Recall that by Lemma~\ref{lem:time}, the expected running time 
of our algorithm is 
$O(n+\sum_i H^V_i)$.
The following is the main theorem of this section.
\medskip
\begin{theorem} \label{thm:main-entropy}
For $H^V_i$, the
entropy of the triangle $\tsearch_i^V$ of $T(V)$ containing $x_i$,
and $H(T(I))$, the entropy of the Delaunay triangulation of $I$,
considered as a labeled graph, 
\[ 
\sum_i H^V_i =  O(n + H(T(I))).
\]
\end{theorem}

\begin{proof}
Let $\tsearch^V \eqdef (\tsearch_1^V, \ldots, \tsearch_n^V)$ be 
the vector of all 
the triangles that contain the $x_i$'s. By Claim~\ref{clm:joint}, we have
$H(\tsearch^V) = \sum_i H_i^Y$. Now we apply 
Lemma~\ref{lem:relate-entropy} with
$\cU = \left(\R^2\right)^n$, $X = T(I)$ 
and $Y$. In Lemma~\ref{lem:computeconflict} we will show
that the function $f: (I, T(I)) \mapsto (\tsearch^V_1, \ldots, \tsearch^V_n)$ 
can be computed
in linear time, so $H(\tsearch^V_i) = O(n + H(T(I))$, by 
Lemma~\ref{lem:relate-entropy}. 
This proves the theorem.
\end{proof}
\medskip

We first define some notation --- for a point set 
$P \subseteq V \cup I$ and $p \in P$, 
let $\Gamma_P(p)$ denote the neighbors of $p$ in $T(P)$. It 
remains to prove the following lemma.\footnote{A similar
lemma is used in~\cite{ChazelleMu09} in the context of 
hereditary algorithms for three-dimensional polytopes.}
\medskip

\begin{lemma}\label{lem:computeconflict}
Given $I$ and $T(I)$, for every $x_i$ in $I$ we can compute the 
triangle $\tsearch^V_i$ in $T(V)$ that contains $x_i$ 
in total expected time $O(n)$.
\end{lemma}

\begin{proof}
First, we compute $T(V \cup I)$ from $T(V)$ and $T(I)$ in linear 
time~\cite{C,Kirkpatrick79}. Thus, we now know $T(V \cup I)$ and
$T(V)$, and we want to find for every point $x_i \in I$ the
triangle $\tsearch_i^V$ of $T(V)$ that contains it.
For the moment, let us be a little less ambitious and
try to determine for each $x_i \in I$, a \emph{conflict triangle}
$\tconf_i^V$ in $T(V)$, ie, $\tconf_i^V$ is a triangle $t$ with 
$x_i \in Z_t$.
If $x \in I$ and $v \in V$ such that
$\overline{xv}$ is an edge of $T(V \cup I)$, we can find
a conflict triangle for $x$ in $T(V)$ in 
time $O(n)$ by inspecting all the incident triangles of
$v$ in $T(V)$. Actually, we can find
conflict triangles for \emph{all} neighbors
of $v$ in $T(V \cup I)$ that lie in $I$, by
merging the two neighbor lists (see below).
Noting that on average the size of these lists 
will be constant, we could almost determine
all the $\tconf_i^V$, except for one problem:
there might be inputs $x \in I$ that are not 
adjacent to any $v \in V$ in $T(V \cup I)$.
Thus, we need to dynamically modify $T(V)$
to ensure that there is always a neighbor present.
Details follow.
\begin{claim}\label{clm:neighborfacets}
Let $p \in V \cup I$ and write $V_p \eqdef V \cup \{p\}$. Suppose 
that $T(V \cup I)$ and $T(V_p)$ are known.  
Then,
in total time $O(|\Gamma_{V \cup I}(p)| + |\Gamma_{V_p}(p)|)$,
for every $x_i \in \Gamma_{V \cup I}(p) \setminus V_p$,
we can compute a conflict triangle $\tconf^{V_p}_i$ of $x_i$ in 
$T(V_p)$.
\end{claim}

\begin{proof}
Let $x_i \in \Gamma_{V \cup I}(p) \setminus V_p$,
and let $\tconf^{V_p}_i$ be the triangle of $T(V_p)$ incident to $p$ that
is intersected by line segment $\overline{px_i}$. We claim that 
$\tconf^{V_p}_i$ is
a conflict triangle for $x_i$. Indeed, since $\overline{px_i}$ is an edge
of $T(V \cup I)$, by the characterization of Delaunay edges 
(eg,~\cite[Theorem~9.6(ii)]{deBergKrOvSc00}), there exists an circle 
$C$ through $p$ 
and $x_i$ which does not contain any other points from 
$V \cup I$. In particular, $C$ does not contain any other
points from $V_p \cup \{x_i\}$. Hence $\overline{px_i}$ is also an edge
of $T(V_p \cup \{x_i\})$, again by the characterization of Delaunay
edges applied in the other direction. Therefore, triangle 
$\tconf^{V_p}_i$ is
destroyed when $x_i$ is inserted into $T(V \cup J)$, and is
a conflict triangle for $x_i$ in $T(V_p)$.
It follows that the conflict triangles for 
$\Gamma_{V \cup I}(p) \setminus V_p$ 
can be
computed by merging the cyclically ordered lists $\Gamma_{V \cup I}(p)$ and 
$\Gamma_{V_p}(p)$. This requires a number of steps that is linear of 
the size of the two lists, as claimed.
%\qed
\end{proof}

For certain pairs of points $p,x_i$, the previous claim provides
a conflict triangle $\tconf_i^{V_p}$. The next claim allows us
to get $\tconf_i^V$ from this, which is what we wanted
in the first place.

\begin{claim}\label{clm:conflict}
Let $x_i \in I$ and let $p \in V \cup I$. Let $\tconf_i^{V_p}$ be the conflict
triangle for $x_i$ in $T(V_p)$ incident to $p$, as determined in 
Step~\ref{step:applyClaim}. Then we can find
a conflict triangle $\tconf_i^V$ for $x_i$ in $T(V)$ in constant time.
\end{claim}

\begin{proof}
If $p \in V$, there is nothing to prove, so assume that $p \in I$.
If $\tconf_i^{V_p}$ has all vertices in $V$, then it is also
a triangle in $T(V)$, and we are trivially done.
So assume that one vertex of $\tconf_i^{V_p}$ is $p$.
Let $e$ be the edge of $\tconf_i^{V_p}$ not incident to $p$, and let
$v,w$ be the endpoints of $e$.
We will show that $x_i$ is in conflict with at least one of the two
triangles in $T(V)$ that are incident to $e$. Given $e$, such a triangle can 
clearly be found in constant time. Refer to Fig.~\ref{fig:conflict} 
for a depiction of the following arguments.

Since $v,w \in V$, by the characterization
of Delaunay edges, it follows that $e$ is also an edge of $T(V)$. 
If $x_i$ does not lie in $\tconf_i^{V_p}$, then $x_i$ must 
also be in conflict with the other triangle $t$ that is incident
to $e$ (since $t$ is intersected by the Delaunay edge $\overline{px_i}$).
Note that $t$ cannot have $p$ as a vertex and is a triangle of $T(V)$.

Suppose $x_i$ lies in $\tconf_i^{V_p}$. Since $\tconf_i^{V_p}$
is a triangle in $T(V_p)$, the interior has no points other
than $x_i$.
Thus, the segments $\overline{vx_i}$ and $\overline{wx_i}$ 
are edges of $T(V_p \cup \{x_i\})$.
These must also be edges of 
$T(V \cup \{x_i\})$. But this means that $x_i$ must conflict
with the triangle in $T(V)$ incident to $e$ at the same side as
$\tconf_i^{V_p}$.
\end{proof}
\medskip

\begin{algorithm}
\begin{enumerate}
\item Let $Q$ be a queue containing the elements in $V$. 

\item\label{step:Qloop} While $Q \not= \emptyset$.

\begin{enumerate}
 \item Let $p$ be the next point in $Q$.
 \item\label{step:insert2} If $p = x_i \in I$, then insert $p$ into $T(V)$ 
 using the conflict triangle $\tconf^V_i$ for $x_i$, to obtain $T(V_p)$.
 If $p \in V$, then $T(V_p) = T(V)$.
 \item\label{step:applyClaim} Using Claim~\ref{clm:neighborfacets}, for 
            each unvisited neighbor $x_j \in \Gamma_{V \cup I}(p) \cap I$, 
	    compute a conflict triangle $\tconf^{V_p}_j$ in $T(V_p)$.
 \item\label{step:getSfacet} For each unvisited neighbor 
           $x_j \in \Gamma_{V \cup I}(p) \cap I$, using
           $\tconf^{V_p}_j$, compute a conflict triangle $\tconf^V_j$ 
	   of $x_j$ in $T(V)$. Then
           insert $x_j$ into $Q$, and mark it as visited.
\end{enumerate}
\end{enumerate}
\caption{Determining the conflict triangles.}
\label{alg:conflicts}
\end{algorithm}

The conflict triangles for all points in $I$ can now be computed using 
breadth-first search (see Algorithm~\ref{alg:conflicts}). 
The loop in Step~\ref{step:Qloop} maintains the invariant that for each 
point $x_i \in Q \cap I$, a conflict triangle $\tconf^V_i$ in $T(V)$ is known. 
Step~\ref{step:insert2} is performed as in the traditional randomized
incremental construction of Delaunay 
triangulations~\cite[Chapter~9]{deBergKrOvSc00}: walk from $\tconf_i^V$
through the dual graph if $T(V)$ to determine the conflict set $S_i$
of $x_i$ (as in the proof of Claim~\ref{clm:getting-si}), insert new edges
from all points incident to the triangles in $S_i$ to $x_i$, and remove
all the old edges that are intersected by these new edges. The properties
of the conflict set ensure that this yields a valid Delaunay triangulation.
By Claim~\ref{clm:conflict}, Step~\ref{step:getSfacet} can be performed
in constant time.

\begin{figure}[t]
\begin{center}
\includegraphics{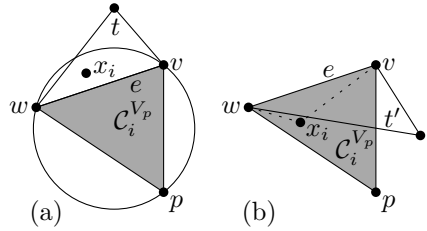}
\end{center}
\caption{(a) If $x_i$ is outside $\tconf_i^{V_p}$, it conflicts
with the triangle $t$ of $T(V)$ on the other side of $e$.
(b) If $x_i$ lies inside $\tconf_i^{V_p}$, it conflicts with the
triangle $t'$ of $T(V)$ at the same side of $e$, since
$\overline{vx_i}$ and $\overline{wx_i}$ are both
edges of $T(V)$.}\label{fig:conflict}
\end{figure}

\medskip

The loop in Step~\ref{step:Qloop} is executed at most once for each 
$p \in V \cup I$.
It is also executed at least once for each point, since $T(V \cup I)$ is 
connected
and in Step~\ref{step:getSfacet} we perform a BFS. The insertion in 
Step~\ref{step:insert2} takes $O(|\Gamma_{V_{x_i}}(x_i)|)$ time. 
Furthermore,
by Claim~\ref{clm:neighborfacets}, the conflict triangles 
of $p$'s neighbors in $T(V \cup I)$ can be computed in 
$O(|\Gamma_{V_p}(p)| + |\Gamma_{V \cup I}(p)|)$
time. Finally, as we argued above, Step~\ref{step:getSfacet} can be carried 
out in total $O(|\Gamma_{V \cup I}(p)|)$ time. 
Now note that for $x_i \in I$, $|\Gamma_{V_{x_i}}(x_i)|$ 
is proportional to $|S_i|$, the number of triangles in $T(V)$ in conflict 
with $x_i$. 
Hence, the total expected running time is proportional to
\begin{multline*}
\EX \Bigl[\sum_{p \in V \cup I}\left(|\Gamma_{V_p}(p)| + 
|\Gamma_{V \cup I}(p)|\right) \Bigr]\\
=
\EX\Bigl[\sum_{v \in V} |\Gamma_V(v)| + \sum_{i = 1}^n |S_i| + 
\sum_{p \in V \cup I} |\Gamma_{V \cup I}(p)|\Bigr]
 = O(n).
\end{multline*}
Finally, using BFS as in the proof of Claim~\ref{clm:getting-si}, 
given the conflict triangles $\tconf^V_i$, 
the triangles $\tsearch^V_i$ that contain the $x_i$'s 
can be found in $O(n)$ expected time, 
and the result follows.
%\qed
\end{proof}

\subsection{The time-space tradeoff}\label{sec:tradeoff}
We show how to remove the assumption that we have prior knowledge of
the $\D_i$'s (to build the search structures $D_i$) and prove
the time-space tradeoff given in Theorem~\ref{thm:del}. These techniques
are identical to those used in \S \ref{sec:SI-sort}.
For the sake of clarity, we give a detailed explanation for this setting.
Let $\eps \in (0,1)$ be any constant. 
The first $\lceil\log n\rceil$ rounds of
the learning phase are used as in \S\ref{sec:learning} 
to construct the Delaunay
triangulation $T(V)$. 
We first build a standard search structure
$D$ over the triangles of $T(V)$~\cite[Chapter~6]{deBergKrOvSc00}. 
Given a point $x$, we can
find the triangle of $T(V)$ that contains $x$ in $O(\log n)$ time.

The learning phase takes $M = cn^{\eps}$ rounds, for some
large enough constant $c$. 
The main trick is to observe that 
(up to constant factors), the only probabilities that are relevant
are those that are at least $n^{-\eps/3}$. In each round, for each $x_i$,
we record the triangle of $T(V)$ that $x_i$ falls into. Fix $i$, and for any
triangle $t$ of $T(V)$, let $\chi_t$ be the number of times over
the first $M$ rounds that $\tsearch_i^V = t$. 
At the
end of $M$ rounds, we take the set $R_i$ of triangles $t$
with $\chi_t > 0$. 
We remind the reader that $p(t,i)$ is the probability that $x_i$ lies
in triangle $t$. The proof of the following lemma is identical to
the proof of Lemma~\ref{lem:learn}.
\begin{lemma}\label{lem:triangle-chernoff}
Fix $i$. With probability at least $1 - 1/n^3$, for every triangle
$t$ of $T(V)$, if $p(t,i) > n^{-\eps/3}$, then
$Mp(t,i)/2 < \chi_t < 3Mp(t,i)/2$. \qquad \endproof
\end{lemma}

For every triangle $t$  in $R_i$, we estimate $p(t,i)$ as 
$\hat{p}(t,i) = \chi_t/M$,  and we use $\hat{p}(t,i)$ to
build the approximate search structure $D_i$. 
For this, we take the planar subdivision $G_i$ induced by the
triangles in $R_i$, compute the convex hull of $G_i$,
and triangulate the remaining polygonal facets.
Then we use the construction of Arya~\etal~\cite{AMM} 
to build an optimal planar point location structure $D_i$ for $G_i$
according to the distribution $\hat{p}_i$ (the triangles of $G_i$ not
in $R_i$ are assigned probability $0$). This structure $G_i$
has the property that a point in a triangle $t$ with probability $\hat{p}(t,i)$
can be located in $O(\log(1/\hat{p}(t,i)))$ 
steps~\cite[Theorems~1.1 and 1.2]{AMM}. 

The limiting phase
uses these structures to find $\tsearch^V_i$ for every $x_i$: given
$x_i$, we use $D_i$ to search for it. 
If the search does
not terminate in $\log n$ steps or $D_i$ fails to find
$\tsearch_i^V$ (ie,  $\tsearch_i^V \notin R_i$), then we use the
standard search structure, $D$, to find $\tsearch_i^V$. Therefore,
we are guaranteed to find $\tsearch^V_i$ in $O(\log n)$ time.
Clearly, each $D_i$ stores $O(M) = O(n^\eps)$ triangles,
so by the bounds given in~\cite{AMM}, each $D_i$ can be constructed 
with size $O(n^\eps)$ in $O(n^\eps\log n)$ time.
Hence, the total space is bounded by $n^{1+\eps}$ and the
time required to build all the $D_i$'s is $O(n^{1+\eps}\log n)$.

Now we just repeat the argument
given in \S\ref{sec:SI-sort}. 
Instead of doing it through words, we 
write down the expressions (for some variety).
Let $s(t,i)$ denote
the time to search for $x_i$ given that $p(i,t) > n^{-\eps/3}$.
By Lemma~\ref{lem:triangle-chernoff}, we have $\chi_t > Mn^{-\eps/3}/2$,
so $t \in R_i$, for $c$ large enough, and thus
$s(t,i) = O(\log(1/\hat{p}(t,i))) = O(1 - \log p(t,i))$.
Thus,
\begin{multline*}
\sum_{t : p(t,i) > n^{-\eps/3} } p(t,i) s(t,i) =  
O\Bigl(\sum_{t : p(t,i) > n^{-\eps/3}} p(t,i) (1- \log p(t,i)) \Bigr)\\
= 
O\Bigl( 1-\sum_{t : p(t,i) > n^{-\eps/3}} p(t,i) \log p(t,i)\Bigr).
\end{multline*}
We now bound the expected search time for $x_i$.
\begin{multline*}
\sum_t p(t,i) s(t,i) =  
\sum_{t :  p(t,i) \leq n^{-\eps/3}} p(t,i) s(t,i) + 
\sum_{t : p(t,i) > n^{-\eps/3}} p(t,i) s(t,i) \\
=  O\Bigl(
1 +
\sum_{t :  p(t,i) \leq n^{-\eps/3}} p(t,i) \log n - 
\sum_{t : p(t,i) > n^{-\eps/3}} p(t,i) \log p(t,i) 
\Bigr)
\end{multline*}
Noting that for $p(t,i) \leq n^{-\eps/3}$, we have 
$O(\log n) = O(\eps^{-1}\log(1/p(t,i)))$, we get 
\begin{align*}
&\sum_t p(t,i) s(t,i) \\
&=  
 O\Bigl(
1 -
\eps^{-1} \sum_{t :  p(t,i) \leq n^{-\eps/3}} p(t,i)\log p(t,i) - 
\sum_{t : p(t,i) > n^{-\eps/3}} p(t,i) \log p(t,i) 
\Bigr)\\
&= O\Bigl(
1 -
\eps^{-1} \sum_{t} p(t,i)\log p(t,i)  
\Bigr)
= O(1+\eps^{-1}H_i^V).
\end{align*}
If follows that the total expected search time is 
$O(n + \eps^{-1}\sum_i H^V_i)$. By the
analysis of \S\ref{sec:algorithm} and Theorem~\ref{thm:main-entropy},
we have that the expected running time in the limiting phase is 
$O(\eps^{-1}(n + H(T(I))))$.
If the conditions in Lemmas~\ref{lem:learning} 
and~\ref{lem:triangle-chernoff} do not hold, then
the training phase fails. But this happens with probability
at most $1/n$.
This completes the proof of Theorem~\ref{thm:del}.

\section{Conclusions and future work}

Our overall approach has been to deduce a ``typical" instance for the
distribution, and then use the solution for the typical instance to solve the
current problem.  This is a very appealing paradigm - even though the actual
distribution $\D$ could be extremely complicated, it suffices to learn just
\emph{one} instance.  It is very surprising that such a single instance exists
for product distributions.  One possible way of dealing with more general
distributions is to have a small set of typical instances. It seems plausible
that even with two typical instances, we might be able to deal with some
dependencies in the input.

We could imagine distributions that are very far from being generated by
independent sources. Maybe we have a graph labeled with numbers, and the input
is generated by a random walk. Here, there is a large dependency between
various components of the input. This might require a completely different
approach than the current one.

Currently, the problems we have focused upon already have $O(n\log n)$ time
algorithms.  So the best improvement in the running time we can hope for is a
factor of $O(\log n)$. The entropy optimality of our algorithms is extremely
pleasing, but our running times are always between $O(n)$ and $O(n\log n)$.  It
would be very interesting to get self-improving algorithms for problems where
there is a much larger scope for improvement. Ideally, we want a problem where
the optimal (or even best known) algorithms are far from linear. Geometric
range searching seem to a good source of such problems. We are given some set
of points and we want to build data structures that answer various geometric
queries about these points~\cite{AE98}. Suppose the points came from some
distribution.  Can we speed up the construction of these structures? 

A different approach to self-improving algorithms would be to change the input
model. We currently have a memoryless model, where each input is independently
drawn from a fixed distribution.  We could have a Markov model, where the input
$I_k$ depends (probabilistically) only on $I_{k-1}$, or maybe on a small number
of previous inputs.

%For example, a Bayesian version of self-improvement would 
%postulate a prior and treat the $I_k$'s as 
%data conditioning a posterior distribution. 
%One could also consider time-varying distributions or Markov models.
%Of course, a purely adversarial model might easily defeat 
%self-improvement: it would observe how the improvement proceeds
%and render it ineffective by tailoring distributions changing over time.
%Memoryless sources are obviously the place to start any investigation
%on self-improvement. We also believe that 
%the assumption is far less
%restrictive than for online computation.
%Take speech for example. The weakness of a memoryless 
%model is that the next utterance is highly
%correlated with the previous ones: hence the use of Markov models.
%A self-improving algorithm would operate at the level
%of a sentence or a paragraph---not an utterance---where 
%correlations are more diffuse and a memoryless source
%might be a good first approximation.

\bibliographystyle{siam}
\bibliography{journal-selfim}

\appendix
\section{Constructing the $\eps$-net $V$}\label{app:epsnet}
Recall that $\lambda = \lceil \log n \rceil$.
Given a set $\hat I$ of $m \eqdef n\lambda$ points in the plane,
we would like to construct a set $V \subseteq \hat I$ of size
$O(n)$ such that any open disk $C$ with 
$|C \cap \hat I| > \lambda$ intersects $V$. (This is
a $(1/n)$-net for disks.)
We describe how to construct $V$ in deterministic
time $n(\log n)^{O(1)}$, using  a technique by
Pyrga and Ray~\cite{PyrgaRa08}. This is by no means the
only way to obtain $V$. Indeed, it is possible 
to use the older techniques of Clarkson and Varadarajan~\cite{CV}
to get a another---randomized---construction with a better running time.

We set some notation. For a set of points $S$, a \emph{$k$-set}
of $S$ is a subset of $S$ of size $k$ obtained by intersecting $S$
with an open disk. A \emph{$(>k)$-set} is is such a subset with size 
more than $k$.
We give a small sketch of the construction. We take the
collection $\hat I_{=\lambda}$ of  all $\lambda$-sets
of $\hat I$. 
We need to obtain a small hitting set for $\hat I_{=\lambda}$.
To do this, we trim $\hat I_{=\lambda}$ to
a collection of $\lambda$-sets that have small pairwise intersection. 
Within each
such set, we will choose an $\eps$-net (for some  $\eps$). The union
of these $\eps$-nets will be our final $(1/n)$-net.
We now give the algorithmic construction of this set and argue that 
it is a $(1/n)$-net. Then, we will show that it has size $O(n)$.

%Let $\hat I_{= \lambda}$ be the collection of all $\lambda$-sets
%of $\hat I$.
It is well known that the collection
$\hat I_{= \lambda}$ has $O(m \lambda)$ sets~\cite{CS89,Lee82} and
that an explicit description of $\hat I_{= \lambda}$ 
can be found in time $O(m \lambda^2)$~\cite{AggarwalGuSaSh89,Lee82},
since $\hat I_{=\lambda}$ corresponds to the 
$\lambda$-th-order Voronoi diagram of $\hat I$, each of whose
cells represents some $\lambda$-set of $\hat I$~\cite{Lee82}.
Let $\cI \subseteq \hat I_{= \lambda}$ be a maximal subset
of $\hat I_{= \lambda}$ such that for any $J_1, J_2 \in \cI$,
$|J_1 \cap J_2| \leq \lambda/100$. 
We will show in Claim~\ref{clm:con-I} how to
construct $\cI$ in $O(m\lambda^5)$ time.
To construct $V$, take a $(1/200)$-net $V_J$ for each $J \in \cI$,
and set $V \eqdef \bigcup_{J \in \cI} V_J$.\footnote{That is,
$V_J$ is a subset of $J$ such that any open disk that contains more than
$|J|/200$ points from $J$ intersects $V_J$.} It is well known that each
$V_J$ has constant size and can be found in time 
$O(|J|) = O(\lambda)$~\cite[p. 180, Proof~I]{Chazelle00}.
The set $V$ is an $(1/n)$-net for $\hat I$: if an open disk $C$ intersects
$\hat I$ in more than $\lambda$-points, by the maximality of $\cI$, 
it must intersect a set $J \in \cI$ in more than $\lambda/100$ points. 
Now $V$ contains a $(1/200)$-net for $J$ (recall that $|J| = \lambda$),
so  $V$ must meet the disk $C$.
We will argue in Claim~\ref{clm:netsize} that $|V| = O(n)$. 
This completes the proof.
\medskip

\begin{claim} \label{clm:con-I} The set $\cI$ can
be constructed in time $O(m \lambda^5)$.
\end{claim}

\begin{proof} 
We use a simple greedy algorithm.
For each $J \in \hat I_{= \lambda}$, construct the collection
$J_{>\lambda/100}$ of all $(>\lambda/100)$-sets of $J$. The set $J$ has size
$\lambda$, and the total number of disks defined by the points in $J$
is at most $\lambda^3$. Thus, there are at most $\lambda^3$ sets in 
$J_{>\lambda/100}$,
and they can all be found in $O(\lambda^4)$ time.
Since there are at most $O(m\lambda)$ sets $J$ (as we argued earlier),
the total number of $(>\lambda/100)$-sets is $O(m \lambda^4)$, and
they can be obtained in $O(m \lambda^5)$ time. 
Next, perform a radix sort on the multiset 
$\mathcal{J} \eqdef \bigcup_{J \in \hat I_{=\lambda}} J_{>\lambda/100}$.
This again takes time $O(m\lambda^5)$.
Note that for any 
$J_1, J_2 \in \hat I_{=\lambda}$, $|J_1 \cap J_2| > \lambda/100$
precisely if $J_1$ and $J_2$ share some $(> \lambda/100)$-set.
Now $\cI$ is obtained as follows: pick a set $J \in \hat I _{=\lambda}$,
put $J$ into $\cI$, and use the sorted multiset $\mathcal{J}$ to find all 
$J' \in \hat I_{=\lambda}$ that share a $(>\lambda/100)$-set with $J$. Discard
those $J'$ from $\hat I_{=\lambda}$. Iterate until $\hat I_{=\lambda}$ 
is empty. The resulting set $\cI$ has the desired properties.
\end{proof}

\medskip

\begin{claim} \label{clm:netsize} $|V| = O(n)$.
\end{claim}

\begin{proof} The set $V$ is the union of $(1/200)$-nets
for each set $J \in \cI$. Since each net has constant size,
it suffices to prove that $\cI$ has $O(n)$ sets. This follows from
a charging argument due to Pyrga and Ray~\cite[Theorem~12]{PyrgaRa08}.
They show~\cite[Lemma~7]{PyrgaRa08} how to construct a graph 
$G_\cI = (\cI, E_\cI)$
on vertex set $\cI$ with at most $|E_\cI| \leq 24|\cI|$ edges with 
the following property: for $p \in \hat I$, let $\cI_p$ be the set 
of all $J \in \cI$ that contain $p$, and
let $G_p = (\cI_p, E_p)$ be the induced subgraph 
on vertex set $\cI_p$. Then, for all $p$,  $|E_p| \geq |\cI_p|/4-1$.
Thus,
\[
\sum_{p \in \hat I} (|\cI_p|/4 - |E_p|) \leq |\hat I| = m.
\]
Consider the sum $\sum_{p \in \hat I} |\cI_p|$. All sets
in $\cI$ contain exactly $\lambda$ points, so each set
contributes $\lambda$ to the sum. By double counting,
$\sum_{p \in \hat I} |\cI_p|/4 = \lambda|\cI|/4$. 
Furthermore, an edge 
$(J_1, J_2) \in E_\cI$ can
appear in $E_p$ only if $p \in J_1 \cap J_2$, so again
by double-counting,
\[
\sum_{p \in \hat I} |E_p| \leq \lambda |E_\cI| /100 \leq 24 \lambda |\cI|/100.
\]
Hence, 
$m \geq \sum_{p \in \hat I} (|\cI_p|/4 - |E_p|) \geq \lambda|\cI|/100$, and
$|\cI| = O(m/\lambda) = O(n)$.
\end{proof}
\end{document}